\documentclass[twocolumns,10pt]{IEEEtran}

\usepackage{float}
 
\usepackage{amsbsy}
\usepackage{amssymb}
\usepackage{amscd}
\usepackage{amsmath}

\usepackage{cite,enumerate}
\usepackage{tikz}
\usetikzlibrary{shapes}
\usetikzlibrary{arrows}
\usetikzlibrary{patterns}
\usetikzlibrary{arrows.meta}
\usetikzlibrary{decorations} 
\usetikzlibrary{decorations.pathreplacing}
\usetikzlibrary{decorations.pathmorphing}
\usetikzlibrary{shapes,backgrounds,patterns,decorations.text}
\usetikzlibrary{decorations.markings}
 
\usepackage{pgfplots}
\usepackage{pgfgantt}
\usepackage{pdflscape}
\pgfplotsset{compat=newest} 
\pgfplotsset{plot coordinates/math parser=false}

\def\bA{{\boldsymbol{A}}}

\def\bB{{\boldsymbol{B}}}

\def\bC{{\boldsymbol{C}}}

\def\bD{{\boldsymbol{D}}}
\def\bDel{{\boldsymbol{\Delta}}}

\def\be{{\boldsymbol{e}}}
\def\bE{{\boldsymbol{E}}}

\def\bI{{\boldsymbol{I}}}

\def\bK{{\boldsymbol{K}}}
\def\bLam{{\boldsymbol{\Lambda}}}

\def\bL{{\boldsymbol{L}}}

\def\bp{{\boldsymbol{p}}}

\def\bphi{{\boldsymbol{\phi}}}

\def\bPsi{{\boldsymbol{\Psi}}}

\def\bQ{{\boldsymbol{Q}}}

\def\bs{{\boldsymbol{s}}}

\def\bu{{\boldsymbol{u}}}

\def\bv{{\boldsymbol{v}}}

\def\bw{{\boldsymbol{w}}}

\def\bx{{\boldsymbol{x}}}
\def\bX{{\boldsymbol{X}}}

\def\by{{\boldsymbol{y}}}
\def\bY{{\boldsymbol{Y}}}
\def\bz{{\boldsymbol{z}}}
\def\bZ{{\boldsymbol{Z}}}
\def\bzero{{\boldsymbol{0}}}

\def\mG{{{\mathcal{G}}}}

\newcommand{\pp}[1]{{\left( #1 \right)}}
\newcommand{\ppb}[1]{{\left[ #1 \right]}}
\newcommand{\tr}[1]{{\mbox{Tr} \left\{ #1 \right\}}}

\def\mtp{{{\mbox{mod}\ 2 \pi}}}

\DeclareMathOperator*{\argmin}{argmin}

\newcommand{\E}[1]{{ \mathsf{E} \{ #1  \} }}
\newcommand{\Elp}[1]{{ \mathsf{E} \left\{ #1  \right\} }}

\newcommand{\cov}[1]{{ \mathsf{Cov} \{ #1  \} }}
\newcommand{\var}[1]{{ \mathsf{Var} \{ #1  \} }}

\newcommand{\bBom}{\bB_\Omega}
\newcommand{\bLom}{{{\bL_\Omega}}}
\newcommand{\bZom}{\bZ_\Omega}
\newcommand{\bEbom}{\bE_{\bar\Omega}}
\newcommand{\mGom}{{\mG_\Omega}}

\newcommand{\terminal}[2]
  {\draw[thick] (#1,#2) rectangle (#1-.125,#2-.25);
\draw[thick] (#1,#2) -- (#1,#2+.1);
}

\usepackage[color=black,opacity=1]{background}

\SetBgContents{ \begin{minipage}{35cm}{\copyright 2024 IEEE. Personal use
      of this material is permitted. Permission from IEEE must be obtained
       for all other uses, in any current or future media, including
        reprinting/republishing this material for advertising or promotional purposes,
         creating new collective works, for resale or redistribution to servers or lists, 
         or reuse of any copyrighted component of this work in other works. This paper will appear in the IEEE Transactions on Signal Processing, 2024.}\end{minipage}}
\SetBgScale{.5}
\SetBgAngle{0}
\SetBgPosition{current page.south west}
\SetBgHshift{20cm}
\SetBgVshift{2cm}

\title{Massive Synchrony in Distributed Antenna Systems}

\author{Erik G. Larsson \thanks{The author is with Link\"oping
    University, Dept. of Electrical Engineering (ISY), 581 83
    Link\"oping, Sweden. E-mail:
    \texttt{erik.g.larsson@liu.se}.\\ This work was supported by
    ELLIIT, the KAW Foundation, the Swedish Research Council (VR), and
    the REINDEER project of the European Union's Horizon 2020 research
    and innovation program under grant agreement No. 101013425.}}

\begin{document}
 
\maketitle
	
\begin{abstract}
Distributed antennas must be phase-calibrated (phase-synchronized) for
certain operations, such as reciprocity-based joint coherent downlink
beamforming, to work.  We use rigorous signal processing tools to
analyze the accuracy of calibration protocols that are based on
over-the-air measurements between antennas, with a focus on
scalability aspects for large systems. We show that (i) for some
who-measures-on-whom topologies, the errors in the calibration process
are unbounded when the network grows; and (ii) despite that
conclusion, it is optimal -- irrespective of the topology -- to solve
a single calibration problem for the entire system and use the result
everywhere to support the beamforming. The analyses are exemplified by
investigating specific topologies, including lines, rings, and
two-dimensional surfaces.
\end{abstract}	


\section{Introduction}

A distributed antenna system consists of access points that are spread
out geographically and cooperate phase-coherently on wireless
communication or sensing tasks.  This concept is considered a main
technology component of the 6G physical layer, and variations of it
appear under the names \emph{distributed multiple-input
multiple-output (MIMO)} \cite{wang2013spectral}, \emph{network MIMO}
\cite{venkatesan2007network}, \emph{user-centric MIMO}
\cite{demir2021foundations}, \emph{cell-free massive MIMO}
\cite{ngo2017cell}, \emph{RadioWeaves} \cite{van2019radioweaves}, and
\emph{radio stripes} \cite{InterdonatoBNFL2019}.  Each access point
may have a single service antenna, or an array (panel) of antennas.

At any point in time, each [service] antenna, $n$, is associated with
two complex-valued coefficients, $T_n$ and $R_n$, that account for
hardware imperfections and oscillator synchronization errors, and
which multiply the transmitted and received complex-baseband signals.
Herein, we will only be concerned with phase and set, for the $n$th
antenna, $T_n=e^{-jt_n}$ and $R_n=e^{jr_n}$ for some phase values
$t_n$ and $r_n$, defined $\mtp$. The convention with a minus sign on
$t_n$ facilitates an interpretation of $\{t_n,r_n\}$ in terms of
time-delays \cite{larsson2023phase}.  The phase values $\{t_n,r_n\}$
collectively model two separate effects:
\begin{enumerate}
\item[E1.]  Geographically separated access points may have their own
  local oscillators that drive their radio-frequency mixers.  These
  oscillators are imperfect and noisy, and they differ from their
  nominal specifications.  This results in a time-varying shift
  between the oscillator phases at different access points.  Unless
  all oscillators are locked to a common reference using a
  synchronization cable -- which is costly and even infeasible in some
  deployment scenarios -- this phase shift can fluctuate and grow
  quickly.

\item[E2.]  The transmit and receive electronics branch at every
  antenna will have an unknown phase lag that depends on manufacturing
  variations and imperfections in the circuits. This lag varies also
  varies with time, but slowly, once the equipment has warmed up. The
  exact time constant of the variation depends on the actual hardware;
  in experimental work reported in \cite{shepard2012argos}, the phase
  was substantially constant for hours.
\end{enumerate}

\subsection{Calibration Over-the-Air}

Knowledge of certain relations between $\{t_n,r_n\}$ is essential for
the antennas to cooperate phase-coherently.  That calls for
calibration in order to periodically estimate these values, or
appropriate functions thereof.  We will be concerned with calibration
based on \emph{over-the-air measurements} (in situ) between pairs of
antennas.  Within an access point one relies on coupling among the
antennas; between access points, one relies on radio-frequency
propagation.  Calibration within an access point may alternatively be
aided by dedicated internal calibration loops
\cite{bourdoux2003non,benzin2017internal}, but that will be of no
further concern here.

Our analysis herein will be agnostic to the actual origin of the
variations in $\{t_n,r_n\}$, and to whether different antennas are
co-located in an array or geographically separated.  But we emphasize
that in practice, E1 presents a much greater challenge than E2,
especially if the oscillators are not locked to a common reference.
Also, in practice, in a system with geographically separated
multiple-antenna access points it can be advantageous to separate the
problems of calibrating for E1 and E2: calibrate the antennas within
an access point infrequently, and between the access points more
frequently.\footnote{Alternative terms for [phase] calibration between
access points are \emph{phase synchronization} and \emph{phase
alignment}.  Herein, we use the term \emph{calibration}.}

\subsection{Different Types of Calibration}

Distributed antenna systems can operate with different purposes,
requiring different types of calibration \cite{larsson2023phase}. Of
particular interest are {reciprocity (R) calibration}, and {full (F)
  calibration}. Here we give a concise exposition, to provide context.

\subsubsection{Reciprocity (R) calibration}\label{sec:Rcal}
The system is R-calibrated if $\{t_{n}+r_{n} \}$ are known, $\mtp$, up
to a common constant for all $n$.  R-calibration enables
reciprocity-based, joint coherent operation, relying on uplink pilots
for downlink multiuser MIMO beamforming.  Over-the-air methods for
R-calibration use bidirectional measurements between antennas,
essentially measuring $(t_n + r_n) - (t_{n'} + r_{n'}) $ for different
pairs $(n,n')$.  This can be done for both co-located arrays
\cite{Kaltenberger,vieira2017reciprocity,Zetterberg:2011:EIT:1928509.1972710,chen2022hierarchical,shepard2012argos,lee2017calibration,JiangKDLK2015,luo2019massive,papadopoulos2014avalanche,jiang2018framework}
and for distributed antenna systems
\cite{RBPCMMBP:14:TWC,vieralarsson_pimrc,chen2017distributed,kim2022gradual,cao2023experimental,balan2013airsync,rashid2022frequency,ganesan2023beamsyncTWC}.

Importantly, over-the-air R-calibration works without knowing a priori
the propagation delay (coupling) between antennas.  Therefore, there
is no difference in principle between R-calibrating the antennas in a
single co-located array and [jointly] R-calibrating all antennas in a
distributed antenna system -- other than that in the presence of
oscillator drifts, a distributed system would have to be re-calibrated
more often.  (The main effect in this case is E1; for more discussion
of the need for re-calibration in the presence of oscillator drifts,
see \cite{nissel2022correctly,larsson2023phase}.)

It is important to appreciate the distinction between \emph{channel
estimation} and (R)-\emph{calibration}.  Both are required for joint
coherent, reciprocity-based downlink beamforming to work.  While
channel estimation in multiple-antenna systems is well researched
\cite{demir2021foundations}, the calibration problem has received less
attention.  Yet, the latter is a difficult problem.  In time-division
duplexing (TDD) operation, channel estimates can be obtained, to any
desired degree of accuracy, by sending uplink pilots with appropriate
length, power, and reuse patterns.  But over-the-air phase calibration
requires the transmission of specially designed signals \emph{between}
service antennas, which breaks the TDD flow.

\subsubsection{Full (F) calibration}
The system is F-calibrated if $\{r_n-r_{n'}\}$, $\{t_n-t_{n'}\}$ and
$\{r_n-t_{n}\}$ are known, $\mtp$, for all $(n,n')$.  Variations
exist, for F-calibration only on receive or transmit.  F-calibration
is stronger than R-calibration, and enables the use of geometrically
parameterized array models -- facilitating fingerprinting and
directional (in angle) beamforming.  F-calibration implies
R-calibration, but not conversely.

F-calibration can also be performed over-the-air by performing
measurements between antennas
\cite{aumann1989phased,vieira2016receive}, but this requires the
propagation delay between the antennas to be a priori known
\cite{larsson2023phase} -- which can be challenging in a distributed
antenna system.

\subsection{Specialization to R-Calibration}

In the rest of the paper we consider only R-calibration, although some
of the results also apply, mutatis mutandis, to other types of
calibration.  We define, for each antenna $n$, the phase parameter
$\phi_n=t_n+r_n$.  The calibration objective is then to obtain
estimates, $\{ \hat\phi_n \}$, of $\{ \phi_n \}$, from pairwise
measurements between antennas. We call $\{\hat\phi_n-\phi_n\}$ the
[phase] estimation errors.

In what follows, by system \emph{topology} we refer to the graph that
defines who measures on whom -- not to be confounded with how antennas
are interconnected over backhaul, which is unimportant here.

\subsection{Contributions and Preview of the Results}\label{sec:preview}

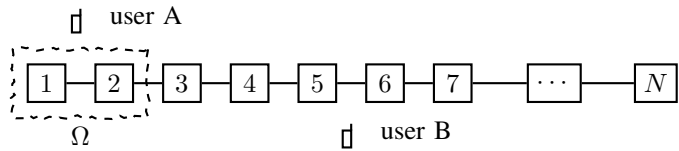
\begin{figure}[t!] 
  \begin{center}
    \begin{tikzpicture}[xscale=.9,yscale=.9]
      \draw[decorate,dashed, thick,decoration={random steps,segment length=1.5pt,amplitude=.8pt}] (1.5,0.5) rectangle (-0.5,-0.5)
               node at (0.5,-0.5) [anchor=north] {$\Omega$};
      
      \node (a1) at (0,0) [draw,rectangle,thick,minimum width=.5cm,minimum height=.5cm] {$1$};
      \node (a2) at (1,0) [draw,rectangle,thick,minimum width=.5cm,minimum height=.5cm] {$2$};
      \node (a3) at (2,0) [draw,rectangle,thick,minimum width=.5cm,minimum height=.5cm] {$3$};
      \node (a4) at (3,0) [draw,rectangle,thick,minimum width=.5cm,minimum height=.5cm] {$4$};
      \node (a5) at (4,0) [draw,rectangle,thick,minimum width=.5cm,minimum height=.5cm] {$5$};
      \node (a6) at (5,0) [draw,rectangle,thick,minimum width=.5cm,minimum height=.5cm] {$6$};
      \node (a7) at (6,0) [draw,rectangle,thick,minimum width=.5cm,minimum height=.5cm] {$7$};
      \node (a10) at (7.5,0) [draw,rectangle,thick,minimum width=.5cm,minimum height=.5cm] {$\cdots$};
      \node (a12) at (9,0) [draw,rectangle,thick,minimum width=.5cm,minimum height=.5cm] {$N$};
      
      \draw [thick] (a1)   -- (a2);
      \draw [thick] (a2)   -- (a3);
      \draw [thick] (a3)   -- (a4);
      \draw [thick] (a4)   -- (a5);
      \draw [thick] (a5)   -- (a6);
      \draw [thick] (a6)   -- (a7);
      \draw [thick] (a7)   -- (a10);
      \draw [thick] (a10)   -- (a12);
      
      \terminal{0.5}{1}
      \terminal{4.5}{-.7}

      \node at (.8,1) [anchor=west] {user A};
      \node at (4.8,-0.7) [anchor=west]  {user B};
      
    \end{tikzpicture}
  \end{center}
  \caption{Line topology, where antennas perform calibration
    measurements on their immediate neighbors.  Two users A and B are
    also shown.\label{fig:lrs}}
\end{figure}

The contribution of this paper is a rigorous analysis of over-the-air
calibration, specifically addressing three questions:
\begin{enumerate}[Q1.]
\item 
  How accurately can $\{ \phi_n \}$ be estimated, for different system
  topologies? Let $N$ be the total number of antennas in the
  system. What happens when $N$ increases: does the calibration
  problem become easier or harder?

  We show that for some topologies, $\var{\hat\phi_n}$ can grow
  unbounded, for all $n$.  This happens, for example, for the line
  (radio stripe) topology in Figure~\ref{fig:lrs}, where antennas
  measure only on their immediate neighbors
  (Section~\ref{sec:stripe}). Note that $\{ \phi_n \}$, of course, are
  bounded as they are only defined $\mtp$, so this asymptotic
  statement means that errors aggregate unfavorably.

\item When performing reciprocity-based beamforming to a user, from a
  \emph{subset} -- to be denoted $\Omega$ -- of the antennas, how
  should calibration be performed? One may either (a) perform
  R-calibration problem for the whole system, or (b) perform
  R-calibration only involving the subset $\Omega$.  For example,
  consider Figure~\ref{fig:lrs}: user A is in the field-of-view of
  antennas 1 and 2; user B is in the field-of-view of antennas 5 and
  6.  When calibrating for beamforming to A, will it be advantageous
  to (a) use calibration measurements among all $N$ antennas, or (b)
  involve only antennas 1 and 2 in the calibration?

  From the answer to Q1, one may intuit that option (b) is preferable
  over (a).  But we prove that (a) is always better: irrespective of
  the topology, it is optimal to solve a single calibration problem
  for the entire system (Section~\ref{sec:gaincomp}).

\item For what topologies is it possible to achieve \emph{massive
synchrony}, such that $\{ \var{\hat \phi_n} \}$ remain bounded, or
  even vanish, when $N\to\infty$?
  
While a complete characterization remains open, we hope that our
analysis offers a starting point (Section~\ref{sec:massivesync}).

\end{enumerate}

\subsection{Related Work}

The most closely related work is the above-cited literature on
R-calibration
\cite{larsson2023phase,shepard2012argos,Kaltenberger,vieira2017reciprocity,Zetterberg:2011:EIT:1928509.1972710,chen2022hierarchical,lee2017calibration,JiangKDLK2015,luo2019massive,RBPCMMBP:14:TWC,vieralarsson_pimrc,chen2017distributed,rashid2022frequency,kim2022gradual,cao2023experimental,balan2013airsync,papadopoulos2014avalanche,jiang2018framework,ganesan2023beamsyncTWC}.
Note that despite its title, \cite{RBPCMMBP:14:TWC} does not consider
the scalability aspects that we analyze herein.

Our results also have some connection to the writings on distributed
synchronization in complex and wireless networks
\cite{simeone2008distributed,wu2010clock,arenas2008synchronization,ghosh2022synchronized,dorfler2014synchronization,lucas2020multiorder}. That
literature, however, deals with protocols for synchronization by
bilateral interaction between devices, whereas we consider calibration
using centrally processed measurements.

\subsection{Notation} 

Lowercase bold symbols are column vectors, and uppercase bold symbols
are matrices. $(\cdot)^*$ represents the complex conjugate.  $\bzero$
is the all-zeros vector/matrix. $\bI$ is the identity matrix, and
$\{\be_n\}$ denote the columns of $\bI$.  For a set $\Omega$,
$\bar\Omega$ denotes its complement.  Given a set of integers,
$\Omega$, $\bE_\Omega$ denotes a matrix comprising the columns
$\{\be_n: n\in\Omega\}$.  $(\cdot)^T$ denotes the transpose of a
vector or matrix, and $(\cdot)^H$ denotes the Hermitian transpose.
$\bu$ is the vector of all ones: $\bu=[1,...,1]^T$.  For a vector
$\bx$, $\Vert\bx\Vert$ is its norm.  For a matrix $\bX$, $\bX^{-1}$ is
its inverse, $\det(\bX)$ is its determinant, and $[\bX]_{kl}$ is its
$(k,l)$ element.  For a positive semidefinite matrix $\bX$,
$\bX^{1/2}$ denotes its positive semidefinite matrix square root. For
symmetric matrices $\bX$ and $\bY$ of compatible dimensions,
$\bX\succcurlyeq\bY$ means that $\bX-\bY$ is positive semidefinite.
	
\section{Estimating $\{\phi_n\}$ from   Measurements}\label{sec:graphmodel}
	
\subsection{Calibration Measurement Model}\label{sec:calmeasmod}
 
All phase values are defined $\mtp$. We will assume that
$|\phi_n-\phi_{n'}|$ is small for all pairs $(n,n')$, so that we can
ignore the $\mtp$ operation when differences between phase values are
of concern. For R-calibration specifically, this means that
$\{t_n+r_n\}$ differ only slightly for different $n$, such that the
calibration only aims at correcting small residual errors; for
example, this is the case if a coarse calibration has been undertaken
previously.

The $N$ antennas perform bidirectional over-the-air calibration
measurements on one another.  Each pairwise interaction, say between
antennas $n$ and $n'$, results in a measurement of $\phi_n-\phi_{n'}$,
$\mtp$.  If antennas $n_m$ and $n'_m$ intercommunicate in the $m$th
measurement, this measurement is,
\begin{align}\label{eq:measeq1}
x_m = \phi_{n_m} - \phi_{{n'_m}} + w_m,
\end{align}
where $ w_m $ is noise.  All differences $\{\phi_n-\phi_{n'}\}$ are
small, and defined $\mtp$; so are $\{x_m\}$.

The calibration objective is to estimate $\{ \phi_1 , ..., \phi_N\}$
from $\{x_1,...,x_M\}$, where $M$ is the number of measurements.  This
problem is ill-posed, irrespective of how large $M$ is or of the
topology: by adding an arbitrary constant to all $\{\phi_n\}$, none of
$\{x_m\}$ changes.  However, with enough measurements, all $\{ \phi_n
\}$ can be estimated up to a common, additive constant.

Throughout, the measurement noise is the only source of randomness,
and all expectations are with respect to this noise. We assume that
the measurement noises $\{ w_m \}$ are zero-mean random variables with
a known positive definite covariance matrix,
\begin{align}
 \bQ = \cov{\bw} ,
\end{align}
where $\bw=[w_1,...,w_M]^T$.

In practice, one may explicitly perform pairwise measurements, as
suggested by (\ref{eq:measeq1}).  Alternatively, one could have each
antenna broadcast a synchronization pilot that is received by several
neighboring antennas simultaneously. In this case, each bidirectional
measurement is completed first when both involved nodes have performed
their broadcast.  For R-calibration of co-located arrays, such
broadcasting strategies were developed in
\cite{jiang2018framework,papadopoulos2014avalanche}.  Hereafter, the
term ``measurement'' means an estimate of $\phi_n-\phi_{n'}$ for an
antenna pair $(n,n')$, irrespective of whether this estimate is
obtained by a single, direct bidirectional measurement, averaging of
multiple bidirectional measurements, or the use of a broadcast scheme.

\subsection{Graph Representation of the Measurement Model}\label{sec:graphrep}

We represent the measurement topology by an undirected graph $\mG$,
whose nodes represent the $N$ antennas and whose edges represent the
$M$ measurements.  Let $\bB$ be the $M\times N$ incidence matrix of
$\mG$, whose $m$th row corresponds to the $m$th edge of $\mG$ (the
$m$th measurement).  For each $m$, $[\bB]_{mn_m}=1$ and
$[\bB]_{mn'_m}=-1$. All other elements in the $m$th row of $\bB$ are
zero.  Furthermore, let ${\bx=[x_1,...,x_M]^T}$ and ${\bphi =
  [\phi_1,...,\phi_N]^T}$. Then
\begin{align}\label{eq:x}
\bx = \bB \bphi + \bw.
\end{align}
We define the pre-whitened measurement vector
\begin{align}\label{eq:xpw}
\bx' = \bQ^{-1/2} \bx = \bQ^{-1/2} \bB \bphi + \bw',
\end{align}
in which the effective noise
\begin{align}\label{eq:wpw}
\bw' = \bQ^{-1/2} \bw
\end{align}
has covariance matrix $\cov{\bw'} = \bI$.

The two cases of interest are:
\begin{enumerate}[(a)]
\item All $N$ antennas are involved in the calibration. In this case,
  we denote the graph by $\mG$ and its incidence matrix by $\bB$, as
  already defined. The measurements are represented by the vector
  $\bx$ in (\ref{eq:x}), or equivalently, the pre-whitened vector
  $\bx'$ in (\ref{eq:xpw}).

\item Only a subset, $\Omega$, of the antennas are involved in the
  calibration.  The corresponding graph is a subgraph of $\mG$, with
  $N$ nodes and $M_\Omega$ edges, where $M_\Omega$ is the number of
  edges between nodes in $\Omega$.  We denote this subgraph by
  $\mGom$, and its $M_\Omega\times N$ incidence matrix by $\bBom$.
  Let $\bx_\Omega$ be the $M_\Omega$-vector comprising the
  measurements among the antennas in $\Omega$; without loss of
  generality, we can assume that the $M_\Omega$ last elements of $\bx$
  contain $\bx_\Omega$, such that $\bx_\Omega=[\bzero\ \ \bI] \bx$.
  We denote the pre-whitened measurements by $\bx'_\Omega =
  \bQ_\Omega^{-1/2} \bx_\Omega$, where
\begin{align}
\bQ_\Omega  & = \begin{bmatrix}   \bzero & \bI \end{bmatrix}  \bQ \begin{bmatrix}   \bzero \\ \bI \end{bmatrix}  
\end{align}
is the $M_\Omega\times M_\Omega$ lower-right corner of $\bQ$, that is,
the part of the noise covariance matrix associated with the $M_\Omega$
measurements.

\end{enumerate}

\begin{figure}[t!]
  \begin{center}
    \begin{tikzpicture}[xscale=.8,yscale=.8]
      \draw[decorate,dashed, thick,decoration={random steps,segment length=1.5pt,amplitude=.8pt}] (1.35,2.35) rectangle (-0.5,-0.5)
          node at (-0.5,1.5) [anchor=east] {$\Omega$};

          \draw[decorate,dashdotted, thick,decoration={random steps,segment length=1.5pt,amplitude=.8pt}] (1.65,2.65) -- (-0.5,2.65)
          -- (-0.5,3.5) -- (3.5,3.5) -- (3.5,-0.5) -- (1.65,-0.5) -- (1.65,2.65) node at (3.5,1.5) [anchor=west] {$\bar\Omega$};
          
          \node (a1) at (0,0) [draw,circle,thick] {};
          \node (a2) at (1,0) [draw,circle,thick] {};
          \node (a3) at (2,0) [draw,circle,thick] {};
          \node (a4) at (3,0) [draw,circle,thick] {};
          \node (a5) at (0,1) [draw,circle,thick] {};
          \node (a6) at (1,1) [draw,circle,thick] {};
          \node (a7) at (2,1) [draw,circle,thick] {};
          \node (a8) at (3,1) [draw,circle,thick] {};
          \node (a9) at (0,2) [draw,circle,thick] {};
          \node (a10) at (1,2) [draw,circle,thick] {};
          \node (a11) at (2,2) [draw,circle,thick] {};
          \node (a12) at (3,2) [draw,circle,thick] {};
          \node (a13) at (0,3) [draw,circle,thick] {};
          \node (a14) at (1,3) [draw,circle,thick] {};
          \node (a15) at (2,3) [draw,circle,thick] {};
          \node (a16) at (3,3) [draw,circle,thick] {};
          
          \draw [thick] (a1)   -- (a2);
          \draw [thick] (a2)   -- (a3);
          \draw [thick] (a3)   -- (a4);
          \draw [thick] (a5)   -- (a6);
          \draw [thick] (a7)   -- (a8);
          \draw [thick] (a9)   -- (a10);
          \draw [thick] (a10)   -- (a11);
          \draw [thick] (a13)   -- (a14);
          \draw [thick] (a15)   -- (a16);
          \draw [thick] (a1)   -- (a5);
          \draw [thick] (a2)   -- (a6);
          \draw [thick] (a4)   -- (a8);
          \draw [thick] (a5)   -- (a9);
          \draw [thick] (a7)   -- (a11);
          \draw [thick] (a8)   -- (a12);
          \draw [thick] (a9)   -- (a13);
          \draw [thick] (a10)   -- (a14);
          \draw [thick] (a11)   -- (a15);
          \draw [thick] (a12)   -- (a16);
          \draw [thick] (a2)   -- (a7);
          \draw [thick] (a7)   -- (a12);
          \draw [thick] (a6)   -- (a11);
          \draw [thick] (a5)   -- (a2);
          \draw [thick] (a15)   -- (a12);
          \draw [thick] (a10)   -- (a7);

          \begin{scope}[shift={(5.5,0)}]
            
            \node (a1) at (0,0) [draw,circle,thick] {};
            \node (a2) at (1,0) [draw,circle,thick] {};
            \node (a3) at (2,0) [draw,circle,thick] {};
            \node (a4) at (3,0) [draw,circle,thick] {};
            \node (a5) at (0,1) [draw,circle,thick] {};
            \node (a6) at (1,1) [draw,circle,thick] {};
            \node (a7) at (2,1) [draw,circle,thick] {};
            \node (a8) at (3,1) [draw,circle,thick] {};
            \node (a9) at (0,2) [draw,circle,thick] {};
            \node (a10) at (1,2) [draw,circle,thick] {};
            \node (a11) at (2,2) [draw,circle,thick] {};
            \node (a12) at (3,2) [draw,circle,thick] {};
            \node (a13) at (0,3) [draw,circle,thick] {};
            \node (a14) at (1,3) [draw,circle,thick] {};
            \node (a15) at (2,3) [draw,circle,thick] {};
            \node (a16) at (3,3) [draw,circle,thick] {};
            
            \draw [thick] (a1)   -- (a2);
            \draw [thick] (a5)   -- (a6);
            \draw [thick] (a9)   -- (a10);
            \draw [thick] (a1)   -- (a5);
            \draw [thick] (a2)   -- (a6);
            \draw [thick] (a5)   -- (a9);
            \draw [thick] (a5)   -- (a2);
          \end{scope}
          \draw [ decoration={brace,mirror }, decorate] (-0.5,-0.7) -- (3.5,-0.7) node [pos=0.5,anchor=north,yshift=-1mm] {$\mG$};
          \draw [ decoration={brace,mirror }, decorate] (5,-0.7) -- (9,-0.7) node [pos=0.5,anchor=north,yshift=-1mm] {$\mGom$};
          
    \end{tikzpicture}
  \end{center}
  \caption{Example of  $\mG$, $\mGom$, $\Omega$, and $\bar\Omega$.
    Here $N=16$, $N_\Omega=6$, $M=25$ and $M_\Omega=7$. \label{fig:defomega}}
\end{figure}

We assume that the nominal graph $\mG$ is connected.  The graph
$\mGom$ has $N-N_\Omega$ isolated nodes, corresponding to antennas
that do not participate.  We assume that the part of $\mGom$
corresponding to the $N_\Omega$ antennas that actually participate, is
connected. Figure~\ref{fig:defomega} shows an example.

The Laplacian of the nominal graph, $\mG$, is $\bB^T \bB $.  To
accommodate a general noise covariance matrix we define the matrix
\begin{align}\label{eq:defL}
  \bL  & = \bB^T \bQ^{-1} \bB .
\end{align}
If $\bQ=\bI$, $\bL$ reduces to the standard graph Laplacian. If $\bQ$
is diagonal, $\bL$ becomes the Laplacian of a weighted graph (with
weights equalling the reciprocal measurements variances).  For general
$\bQ$, the matrix $\bL$ is not a conventional Laplacian matrix, but
its nullspace is the same as that of the Laplacian $\bB^T \bB $, and
this is what is required by the subsequent analysis.

Since $\mG$ is connected, the nullspace of $\bB^T \bB $, and therefore
of $\bL$, is one-dimensional and spanned by
$\bu$\cite{godsil2001algebraic}.  More specifically, by a basis
change, $\bL$ can be written as,
\begin{align}\label{eq:Ldecomp}
  \bL  = [\bu\ \bZ] \begin{bmatrix} 0 & \bzero \\ \bzero & \bLam \end{bmatrix} \begin{bmatrix} \bu^T \\ \bZ^T \end{bmatrix},
\end{align}
where $\bLam$ is an $(N-1)\times (N-1)$ diagonal matrix whose
(positive) diagonal elements are the non-zero eigenvalues of $\bL$,
and where $\bZ$ is an $N\times (N-1)$ matrix whose columns are
eigenvectors of $\bL$ and constitute an orthonormal basis for the
orthogonal complement of $\bu$.  Note that (\ref{eq:Ldecomp}) is
similar to the eigenvalue decomposition, but not identical thereto as
$\bu$ does not have unit norm.

The Laplacian of the weighted subgraph, $\mGom$, is given by $\bBom^T
\bBom$. Similarly to above, we define
\begin{align}
  \bLom  & = \bBom^T \bQ_\Omega^{-1} \bBom,
\end{align}
which again, is a (weighted) graph Laplacian if $\bQ$ is diagonal.
The nullspace of $\bLom$ has dimension $N_{\bar\Omega}+1$.  Through a
basis change, we can write,
\begin{align}\label{eq:Lomdecomp}
  \bLom  = [\bu\ \bEbom\  \bZom] 
  \begin{bmatrix} 
    0 & \bzero & \bzero  \\ 
    \bzero & \bzero & \bzero  \\ 
    \bzero & \bzero & \bLam_\Omega 
  \end{bmatrix} 
  \begin{bmatrix} \bu^T \\ \bEbom^T \\ \bZom^T  
  \end{bmatrix},
\end{align}
for some diagonal $\bLam_\Omega$, where $\bZom$ is an $N\times
(N_\Omega-1)$-dimensional matrix whose columns constitute an orthonormal basis
of the orthogonal complement of $\{\bu,\bEbom\}$.  Note that
(\ref{eq:Lomdecomp}) is not the eigenvalue decomposition, since (among
others) $\bu$ and the columns of $\bEbom$ are not orthogonal.

For future use, we note the following facts:
\begin{itemize} 
\item For $\bZ$, it holds that
  \begin{align}
    \bZ^T\bu  & = \bzero, \\
    \bZ^T\bZ & =\bI, \\
    \bZ\bZ^T  & =\bI - \frac{1}{N} \bu\bu^T  \label{eq:Zproj} \\
    \bZ^T\bL\bZ & = \bLam. \label{eq:ZLZL}
  \end{align}
  
\item For $\bZom$, it holds that
  \begin{align}
    \bZom^T\bu  & = \bzero, \\
    \bZom^T\be_n & =\bzero, \qquad n\notin\Omega \\
    \bZom^T\bZom & = \bI, \\
    \bZom\bZom^T  & =\bI - 
       [ \bu \ \   \bEbom ]
        \pp{ \begin{bmatrix} \bu^T \\ \bEbom^T \end{bmatrix}  [ \bu \ \   \bEbom ]  }^{-1}
        \begin{bmatrix} \bu^T \\ \bEbom^T \end{bmatrix} .      \label{eq:Rproj}
  \end{align}

\item Since $\mG$ is connected, we must have $M\ge N-1$. The nullspace
  of $\bB$ is one-dimensional and spanned by $\bu$:
  $\bB\bu=\bzero$. The column rank of $\bB$ is $N-1$. The columns of
  $\bZ$ span the columns of $\bB^T$ and of $(\bQ^{-1/2}\bB)^T$.

\item Since the ``$\Omega$-part'' of $\mGom$ is connected, we must
  have $M_\Omega \ge N_\Omega-1$. Furthermore, $\bBom$ has column rank
  $N_\Omega-1$. Its nullspace is spanned by $\{\bu,\bEbom\}$. The
  columns of $\bZom$ span the columns of $\bBom^T$ and of
  $(\bQ_\Omega^{-1/2} \bBom)^T$.
\end{itemize}

\subsection{Least-Squares Estimate of $\{\phi_n\}$ in Case (a)}\label{sec:LSa}

In case (a), the least-squares estimate of $\bphi$, given the
pre-whitened measurements in (\ref{eq:xpw}) obtained from the nominal
graph $\mG$, is
\begin{align}\label{eq:lsx}
  \argmin_\bphi \Vert \bx' - \bQ^{-1/2} \bB \bphi \Vert^2.
\end{align}
Since the column space of $(\bQ^{-1/2} \bB)^T$ is spanned by $\bZ$,
$\bphi$ can be identified up to a vector proportional to $\bu$.  More
precisely, (\ref{eq:lsx}) has the solutions
\begin{align}\label{eq:phi}
\hat\bphi =  \bZ\hat\bs + \lambda \bu,
\end{align}
where 
\begin{align}
  \hat \bs = (\bZ^T \bB^T \bQ^{-1} \bB \bZ)^{-1} \bZ^T \bB^T \bQ^{-1} \bx,
\end{align}
and $\lambda$ is an indeterminate scalar.

While $\hat\bs$ is uniquely determined, $\lambda$ is not. Fortunately,
$\lambda$ does not affect beamforming performance
(Section~\ref{sec:main}). Hence, we can set $\lambda=0$.  This yields
the unique solution
\begin{align}\label{eq:hatphi}
  \hat \bphi =  \bZ\hat\bs  = \bZ (\bZ^T \bB^T \bQ^{-1} \bB \bZ)^{-1} \bZ^T \bB^T \bQ^{-1} \bx.
\end{align}
Note, in passing, that among all   solutions to   (\ref{eq:lsx}), (\ref{eq:hatphi}) is the one with the smallest  $\Vert\bphi\Vert$.

A direct calculation yields
\begin{align}
  \E{\hat \bphi}  & =  \E{\bZ (\bZ^T \bB^T \bQ^{-1} \bB \bZ)^{-1} \bZ^T \bB^T \bQ^{-1} (\bB\bphi + \bw) } \nonumber    \\
  & = \bZ (\bZ^T \bB^T \bQ^{-1} \bB \bZ)^{-1} \bZ^T \bB^T \bQ^{-1}  \nonumber \\ & \qquad \cdot
  \Elp{ {\bB  \pp{\bZ\bZ^T+ \frac{1}{N}\bu\bu^T}\bphi + \bw} } \nonumber \\
  & = \bZ\bZ^T\bphi,   \label{eq:ehatphi} \\
  \cov{  \hat \bphi} & = \bZ (\bZ^T \bB^T \bQ^{-1} \bB \bZ)^{-1} \bZ^T
  = \bZ (\bZ^T \bL \bZ)^{-1} \bZ^T . \label{eq:covphi}
\end{align}
In the last step of (\ref{eq:ehatphi}), we used that $\bB\bu=\bzero$
and that $\E{\bw}=\bzero$.  In (\ref{eq:covphi}), we additionally used
that $\cov{\bw}=\bQ$.  Equation (\ref{eq:ehatphi}) implies that the
part of $\hat\bphi$ that lies outside the unidentifiable (and
uninteresting) subspace spanned by $\bu$ is unbiased.  This means that
(\ref{eq:covphi}) completely quantifies the accuracy of $\hat\bphi$.
Note that $\cov{ \hat \bphi}$ is rank-deficient -- a consequence of
the non-identifiability of $\lambda$.

\subsection{Monotonicity of $\cov{\hat\bphi}$ in Case (a)}\label{sec:mono}

For fixed $N$, by adding more measurements the accuracy
of $\{ \hat \bphi\}$ improves.  To see this, suppose the
nominal $\bB$ is replaced by
\begin{align}
  \bB'' = \begin{bmatrix} \bB' \\ \bB \end{bmatrix},
\end{align}
for some $\bB'$ that represents the additional measurements.  Let
$\bQ''$ be the noise covariance of the augmented measurement set, such
that $\bQ$ is the lower-right $N\times N$ submatrix of
$\bQ''$. Explicitly, partition $\bQ''$ according to
\begin{align}
  \bQ'' = \begin{bmatrix} \bar\bQ & \tilde\bQ^T \\ \tilde\bQ & \bQ  \end{bmatrix}.
\end{align}
Let 
\begin{align}
  \bL''=\bB''^T  \bQ''^{-1} \bB'' , 
\end{align}
and let $\bZ''$ be the counterpart of $\bZ$ associated with $\bL''$.
\begin{figure*}
\begin{align}
  \bL'' - \bL & = \bB''^T  \bQ''^{-1} \bB''   - \bB^T \bQ^{-1} \bB \nonumber \\
  & =  \bB''^T  \bQ''^{-1} \bB''  - \bB''^T \begin{bmatrix}     \bzero \\ \bI \end{bmatrix} \bQ^{-1} \begin{bmatrix}    \bzero & \bI \end{bmatrix} \bB'' \nonumber \\
  & =  \bB''^T  \pp{ \bQ''^{-1} - \begin{bmatrix}    \bzero \\ \bI \end{bmatrix} \bQ^{-1} \begin{bmatrix}   \bzero & \bI \end{bmatrix} } \bB'' \nonumber \\
  & =  \bB''^T  \pp{ 
    \begin{bmatrix}  \bI  & -\bar\bQ^{-1}\tilde\bQ^T \\ \bzero & \bI   \end{bmatrix}
    \begin{bmatrix}   \bar\bQ^{-1} & \bzero \\  \bzero & (\bQ - \tilde \bQ \bar\bQ^{-1} \tilde\bQ^T)^{-1}   \end{bmatrix}
    \begin{bmatrix}  \bI  & \bzero \\ -\tilde\bQ\bar\bQ^{-1} & \bI   \end{bmatrix}
    - \begin{bmatrix}   \bzero \\ \bI \end{bmatrix} \bQ^{-1} \begin{bmatrix}  \bzero & \bI  \end{bmatrix} } \bB''  \nonumber \\
  & =  \bB''^T  { 
    \begin{bmatrix}  \bI  & -\bar\bQ^{-1}\tilde\bQ^T \\ \bzero & \bI   \end{bmatrix}
    \begin{bmatrix}   \bar\bQ^{-1}  & \bzero \\  \bzero & (\bQ - \tilde \bQ \bar\bQ^{-1} \tilde\bQ^T)^{-1} - \bQ^{-1}   \end{bmatrix}
    \begin{bmatrix}  \bI  & \bzero \\ -\tilde\bQ\bar\bQ^{-1} & \bI   \end{bmatrix}
  } \bB''  \label{eq:LppmL}
\end{align}
\hrule
\end{figure*}
By rewriting $\bQ''^{-1}$ using a Schur complement, we obtain
(\ref{eq:LppmL}), shown on top of the next page.  Since $ \bQ \succcurlyeq
\bQ - \tilde \bQ \bar\bQ^{-1} \tilde\bQ^T$, we have using Corollary
7.7.4(a) of \cite{matrixanalysis} that
\begin{align}
  (\bQ - \tilde \bQ \bar\bQ^{-1} \tilde\bQ^T)^{-1} - \bQ^{-1} \succcurlyeq \bzero .
\end{align}
Therefore, the matrix in the middle of the right hand side of
(\ref{eq:LppmL}) is positive semidefinite, from which it follows that
the right hand side of (\ref{eq:LppmL}) is positive semidefinite (note
its Gramian form). We conclude that ${\bL''\succcurlyeq\bL}$,
wherefrom it follows that ${\bZ''^T\bL''\bZ''\succcurlyeq
  \bZ''^T\bL\bZ''} $.  By applying again Corollary 7.7.4(a) of
\cite{matrixanalysis}, we find that
\begin{align}\label{eq:ZLZineq}
  (\bZ''^T\bL\bZ'')^{-1} \succcurlyeq (\bZ''^T\bL''\bZ'')^{-1}.
\end{align}

The nullspaces of $\bL$ and $\bL''$ are one-dimensional and spanned by
$\bu$.  Therefore, $\bZ$ and $\bZ''$ have the same column space.  It
follows that there exists an $(N-1)\times(N-1)$ matrix $\bPsi$ such
that
\begin{align}
  \bZ''\bPsi & =\bZ \\
  \bPsi^T\bPsi & = \bPsi\bPsi^T  = \bI.
\end{align}
Therefore,
\begin{align}
  &\quad \ \bZ(\bZ^T\bL\bZ)^{-1}  \bZ^T - \bZ''(\bZ''^T\bL''\bZ'')^{-1}\bZ''^T  \nonumber  \\
  & = \bZ''\bPsi(\bPsi^T\bZ''^T\bL\bZ''\bPsi)^{-1} \bPsi^T\bZ''^T \nonumber \\ & \qquad - \bZ''(\bZ''^T\bL''\bZ'')^{-1}\bZ''^T \nonumber \\
  & =  \bZ''(\bZ''^T\bL\bZ'')^{-1} \bZ''^T - \bZ''(\bZ''^T\bL''\bZ'')^{-1}\bZ''^T \nonumber \\ 
  & = \bZ'' \pp{ (\bZ''^T\bL\bZ'')^{-1}   -  (\bZ''^T\bL''\bZ'')^{-1} } \bZ''^T  \succcurlyeq  \bzero,  \label{eq:monotonicity}
\end{align}
where in the last step we used (\ref{eq:ZLZineq}).  That is, the
difference between $\cov{\hat\bphi}$ without the additional
measurements, and $\cov{\hat\bphi}$ with those measurements, is
positive semidefinite.
 
\subsection{Least-Squares Estimate of  $\{\phi_n\}$ in Case (b)}

In case (b), the corresponding least-squares problem is
\begin{align}\label{eq:lsx2}
  \argmin_\bphi \Vert \bx'_\Omega - \bQ_\Omega^{-1/2} \bBom \bphi \Vert^2.
\end{align}
Problem (\ref{eq:lsx2}) has the solutions
\begin{align}\label{eq:phi2}
  \hat\bphi =  \bZom\hat\bs_\Omega + \lambda \bu + \sum_{n\notin \Omega} \lambda_n \be_n,
\end{align}
where 
\begin{align}
  \hat \bs_\Omega = (\bZom^T \bBom^T \bQ_\Omega^{-1} \bBom \bZom)^{-1} \bZom^T \bBom^T \bQ_\Omega^{-1} \bx_\Omega,
\end{align}
and $\lambda$ and $\{\lambda_n\}$ are indeterminate scalars that we
will set to zero.  The mean and covariance can be found via a similar
calculation as in Section~\ref{sec:LSa}, using the above-established
facts about $\bLom$, $\bZ_\Omega$, and $\bBom$:
\begin{align}
  \E{\hat \bphi}  & = \bZom\bZom^T\bphi ,   \label{eq:ehatphi2} \\
  \cov{  \hat \bphi} & = \bZom (\bZom^T \bLom \bZom)^{-1} \bZom^T . \label{eq:covphi2}
\end{align}

Note that $\var{\hat\phi_n}$ for $n\in\Omega$ is not necessarily
larger in case (b) than in case (a), even though in case (a) we have
more measurements, because in case (a) there are more identifiable
parameters.
 
\section{Beamforming Gain Analysis}\label{sec:main}

Next we examine the impact of estimation errors in $\{\hat\phi_n\}$ on
joint coherent downlink beamforming performance, when targeting a
specific point (focal spot) in space with reciprocity-based
beamforming.  The beamforming is performed by the antennas in the set
$\Omega$.

To keep the analysis clean, we restrict the discussion to the
beamforming of a monochromatic (sinusoidal) signal with some given
carrier frequency.  The argument extends directly to any
signal whose bandwidth is less than the channel coherence bandwidth
(reciprocal excess delay).\footnote{Technically, a finite excess-delay
assumption is inconsistent with an array aperture that grows to
infinity.  In practice, signals originating at antennas farther and
farther away would be attenuated more and more strongly, eventually
contributing negligibly to the channel response. }  Also, nothing
prevents the application of the analysis to multiple narrowband
signals that are adjacent in the frequency domain.  Let $ h_n $ be the
channel frequency response (complex-baseband channel gain) at the
carrier frequency of concern, between the $n$th antenna and the focal
spot, and let $ a_n $ be the respective beamforming weight applied by
the $n$th antenna.

Suppose that the above-described estimates $\{ \hat \phi_n \}$ are
used to pre-compensate $\{a_n \}$ when performing the beamforming.
The effective channel to the focal spot is then
\begin{align}\label{eq:effgfocal}
  g = \sum_{n\in\Omega} h_n a_n e^{j\hat\phi_n} e^{-j \phi_n}.
\end{align}
The corresponding beamforming (power) gain is $|g|^2$. Adding a common
constant to all $\{ \hat \phi_n \}$ does not affect $|g|$, which is why we
can safely set $\lambda=0$ and $\lambda_n=0$ in
Section~\ref{sec:graphmodel}.

\subsection{Beamforming for Constructive Interference}

If $\{ \phi_n \}$ are perfectly known, then $|g|$ can be made to scale
with $N$, by taking the angle of $a_n$ to align with that of $h_n^*$,
causing constructive interference at the focal spot. This gives the
standard coherent array gain.

If $\{ \phi_n \}$ are only imperfectly known, the array gain
deteriorates.  In practice, {accurate} knowledge is not critical.  For
example, suppose $|h_n|=1$ and that $\{\phi_n\}$ for half of the
antennas are perfectly known, but that $\{\phi_n\}$ for the other
antennas are off by $\delta$ radians. The relative loss in $|g|^2$ is
then $|1+1|^2/|1+e^{j\delta}|^2$.  Even if $\delta=\pi/2$ (90$^\circ$)
the loss is only 3 dB.  But beyond 90$^\circ$ errors, the array gain
quickly evaporates. We omit a detailed analysis and instead focus on
the more interesting case of null-steering, next.

\subsection{Beamforming for Destructive Interference}

With null-steering, $\{ a_n \}$ are selected such that the signals
from different antennas interfere \emph{destructively} at the focal
spot:
\begin{align}\label{eq:hnan}
  \sum_{n\in\Omega} h_n a_n   = 0,
\end{align}
attempting to make $g=0$. This is the operational principle of
zero-forcing beamforming for multiuser MIMO
\cite{gesbert2007shifting,bjornson2014optimal}.

To analyze the beamforming gain in the presence of phase estimation
errors, consider first case (a).  Recall that
${\{\hat\phi_n-\phi_n\}}$ have nonzero mean because of the
non-identifiability of the estimation problem; see (\ref{eq:ehatphi}).
For analysis purposes, it will prove useful to introduce the following
intermediate quantities:
\begin{align}
 \tilde\phi_n & = \hat\phi_n - \phi_n + \bar \phi,  \label{eq:deftildephi}
\end{align}
where we defined the average of the phase values,
\begin{align}
  \bar\phi & = \frac{1}{N} \sum_{n=1}^N \phi_n = \frac{\bu^T\bphi}{N} .  \label{eq:defphibar1}
\end{align}
One can think of $\{ \tilde\phi_n \}$ as the ``zero-mean part'' of the
estimation errors, because $\E{\tilde\phi_n}=0$ for all $n$. To see
why this is so, let $\tilde\bphi=[\tilde\phi_1,...,\tilde\phi_N]^T$
and note that from (\ref{eq:Zproj}), (\ref{eq:ehatphi}),
(\ref{eq:deftildephi}) and (\ref{eq:defphibar1})  we have that
\begin{align}
  \E{ \tilde \bphi }    = \bZ\bZ^T \bphi - \bphi + \frac{\bu\bu^T}{N} \bphi = \bzero.
\end{align}
We now re-express the effective channel in (\ref{eq:effgfocal}) in
terms of $\{ \bar\phi_n \}$ and $\{ \tilde\phi_n \}$:
\begin{align}
  g & =    \sum_{n\in\Omega} h_n a_n e^{j\hat\phi_n} e^{-j   \phi_n}    
  =   e^{-j\bar\phi} \sum_{n\in\Omega} h_n a_n e^{j \tilde \phi_n } \nonumber \\
  & \approx e^{-j\bar\phi} \sum_{n\in\Omega} h_n a_n (1+ j \tilde \phi_n) \nonumber  \\
  & = e^{-j(\bar\phi-\pi/2)} \sum_{n\in\Omega} h_n a_n    \tilde \phi_n, \label{eq:approxcalc}
\end{align}
where in the third step we performed a first-order Taylor expansion,
$e^{\times}\approx 1+\times$.  This expansion is justified since
$\{\tilde\phi_n\}$ have zero mean, and in the limit of weak
measurement noise they would fluctuate only slightly.  In the last
step of (\ref{eq:approxcalc}) we used (\ref{eq:hnan}).

Next, consider case (b). The analysis is analogous to the one for case (a),
but with $\bar\phi$ re-defined. Specifically, in contrast to
(\ref{eq:defphibar1}), now we set
\begin{align}
  \bar\phi  &=   \begin{bmatrix} 1 &  \bzero^T \end{bmatrix}
  \pp{ \begin{bmatrix} \bu^T \\ \bEbom^T \end{bmatrix}  [ \bu \ \   \bEbom ]     }^{-1}
  \begin{bmatrix} \bu^T \\ \bEbom^T \end{bmatrix} \bphi ,
  \label{eq:barphi2}
\end{align}
and note that for $n\in\Omega$, it holds that ${ \be_n^T [ \bu
    \ \ \bEbom ] = [ 1 \ \ \bzero^T ] }$; therefore, for $n\in\Omega$,
we have
\begin{align}
  \be_n^T [ \bu \ \   \bEbom ]
  \pp{  \begin{bmatrix} \bu^T \\ \bEbom^T \end{bmatrix}  [ \bu \ \   \bEbom ]   }^{-1}
  \begin{bmatrix} \bu^T \\ \bEbom^T \end{bmatrix} \bphi  = \bar\phi.
  \label{eq:barphi22}
\end{align}
Using (\ref{eq:Rproj}) and (\ref{eq:ehatphi2})  we then find that for $n\in\Omega$,
\begin{align}
\E{\hat\phi_n} & = \be_n^T \bZom \bZom^T \bphi  = \phi_n - \bar\phi .
\end{align}
Letting, as before, $ \tilde\phi_n = \hat\phi_n - \phi_n + \bar \phi, $
we have that
\begin{align}
\E{\tilde\phi_n} = \phi_n - \bar\phi - \phi_n + \bar\phi  = 0
\end{align}
for $n\in\Omega$.  This means that the formula for the effective
channel, (\ref{eq:approxcalc}), applies to case (b), as well.

Consequently, we have for both cases (a) and (b), that
\begin{align}
  g \approx    e^{-j(\bar\phi-\pi/2)} \cdot  \bv^T \tilde\bphi,
\end{align}
where $\bv$ is an $N$-vector whose components are
\begin{align}\label{eq:defv}
v_n = \begin{cases} 
h_n a_n ,& n\in \Omega \\
0, & \mbox{otherwise}.
\end{cases}
\end{align}
In either case, $\tilde\bphi$ has zero mean, and covariance
${\cov{\tilde\bphi}=\cov{\hat\bphi}}$, given by (\ref{eq:covphi}) and
(\ref{eq:covphi2}), respectively.  It follows that
\begin{align}
  \E{g} & \approx 0 , \\
  \var{g} & \approx   \bv^T \cov{\hat \bphi} \bv^* = \bv^H \cov{\hat \bphi} \bv, \label{eq:gvar}
\end{align}
with the respective (real-valued) covariance matrices in
(\ref{eq:covphi}) and (\ref{eq:covphi2}) inserted for $\cov{\hat \bphi}$.

\section{Beamforming Gain Comparison}\label{sec:gaincomp}

Henceforth, we are only interested in null-steering.  The variance of
$g$, $\var{g}$, quantifies the beamforming accuracy in terms of how
much power inadvertently reaches the focal point, for an arbitrary
choice of beamforming weights $\{a_n\}$ satisfying (\ref{eq:hnan}).
(The smaller the variance, the better the accuracy.) We will now
compare this variance, for given $\{a_n\}$, between the following two
cases:
\begin{enumerate}[(a)]
\item Beamforming is undertaken by the subset $\Omega$ of the
  antennas, and $\{\hat \phi_n\}$ are obtained from calibration
  measurements among \emph{all} $N$ antennas, that is, from $\mG$.

\item Beamforming is undertaken by the subset $\Omega$ of the
  antennas, but $\{ \hat \phi_n\}$ are obtained \emph{only} from
  calibration measurements among antennas that participate in the
  beamforming, that is, from $\mGom$.
\end{enumerate}

First note from (\ref{eq:defv}) that (\ref{eq:hnan}) can be
equivalently written as $\sum_{n} v_n=0$; that is, using vector
notation,
\begin{align}
  \bu^T\bv &  =0  .  \label{eq:uv}
\end{align}
Also, note that the constraint that $v_n=0$ for $n\notin\Omega$ in
(\ref{eq:defv}) can be expressed as
\begin{align}
 \be_n^T\bv & =0, \qquad n\notin \Omega . \label{eq:ev}
\end{align}
Taken together, this means that the set of possible beamforming
weights $\{ a_n \}$ is defined by the set of $N$-vectors $\bv$ that
satisfy (\ref{eq:uv}) and (\ref{eq:ev}).  Since the columns of $\bZom$
span the orthogonal complement of the space spanned by
$\{\bu,\bEbom\}$, we know that for an arbitrary vector $\bv$ that
satisfies (\ref{eq:uv})--(\ref{eq:ev}), there exists a unique
$(N_\Omega-1)$-vector $\bp$ such that
\begin{align}
  \bv = \bZom\bp .
\end{align}
In case (a), the variance (\ref{eq:gvar}) is given by the quadratic form
\begin{align}
  V_a & = \bv^H  \bZ  (\bZ^T \bL \bZ)^{-1} \bZ^T  \bv   \nonumber \\
  & = \bp^H  \bZom^T \bZ  (\bZ^T \bL \bZ)^{-1} \bZ^T  \bZom \bp
  \nonumber \\
  & = \bp^H \bK_a \bp,  \label{eq:wcvara}
\end{align}
where we defined the  kernel,
\begin{align}
  \bK_a & = \bZom^T \bZ (\bZ^T \bL  \bZ)^{-1}  \bZ^T \bZom . \label{eq:defKa}
\end{align}
In case (b), the  variance is
\begin{align}
  V_b & = \bv^H  \bZom  (\bZom^T \bLom \bZom)^{-1} \bZom^T  \bv   \nonumber \\
  & =    \bp^H  \bZom^T \bZom  (\bZom^T \bLom \bZom)^{-1} \bZom^T  \bZom \bp    \nonumber \\ 
  & =    \bp^H    (\bZom^T \bLom \bZom)^{-1}  \bp  \nonumber \\ 
  & =    \bp^H   \bK_b \bp,
  \label{eq:wcvarb}
\end{align}
where
\begin{align}
  \bK_b & =  (\bZom^T \bLom \bZom)^{-1}.\label{eq:defKb}
\end{align}

We are now going to establish that $V_b \ge V_a$ for any $\bp$, from
which it then follows that $V_b \ge V_a$ for any $\bv$ that satisfies
(\ref{eq:uv})--(\ref{eq:ev}), and therefore for any $\{ a_n \}$ that
satisfy (\ref{eq:hnan}).  Clearly, this is the case if we can
demonstrate that $\bK_b\succcurlyeq\bK_a$.  From Corollary 7.7.4(a) of
\cite{matrixanalysis}, we know that $\bK_b\succcurlyeq \bK_a$ if and
only if $\bK_a^{-1}\succcurlyeq\bK_b^{-1}$. Therefore, the
sought-after result follows if we can show that
$\bK_a^{-1}-\bK_b^{-1}\succcurlyeq\bzero$.

The complication in the analysis lies in the rank-deficiency of $\bL$
and $\bLom$.  To tackle this, note that $\bZ$ and $[\bZom\ \bEbom]$
have the same column space; hence, there exists an $(N-1)\times (N-1)$
matrix $\bPsi$ such that
\begin{align}
  \bZ\bPsi  & = [\bZom\ \ \bEbom],  \label{eq:zpsi} \\
  \bPsi^T\bPsi  & = \bPsi\bPsi^T = \bI.
\end{align}
By multiplying (\ref{eq:zpsi}) from the left by $ \bZom^T$ and from
the right by $\bPsi^T$, we find that
\begin{align}
  \bZom^T\bZ & =    [\bI \ \  \bzero]\bPsi^T.
\end{align}
It follows that  
\begin{align}
  \bK_a & = \bZom^T \bZ (\bZ^T \bL  \bZ)^{-1}  \bZ^T \bZom \nonumber \\
  & = \begin{bmatrix} \bI & \bzero \end{bmatrix}
  \bPsi^T \pp{ \bZ^T \bL \bZ }^{-1} \bPsi
  \begin{bmatrix} \bI \\ \bzero \end{bmatrix} \nonumber \\
  & = \begin{bmatrix} \bI & \bzero \end{bmatrix}
  \pp{ \begin{bmatrix} \bZom^T \\ \bEbom^T \end{bmatrix}
    \bL
    \begin{bmatrix} \bZom & \bEbom \end{bmatrix} }^{-1}
  \begin{bmatrix} \bI \\ \bzero \end{bmatrix} \nonumber \\
  & = \pp{ \bZom^T\bL\bZom - \bZom^T\bL\bEbom (\bEbom^T \bL
    \bEbom)^{-1} \bEbom^T \bL \bZom}^{-1}, \label{eq:Ka2}
\end{align}
where in the last step, we used the block matrix inversion
lemma.\footnote{Whenever the   inverses exist \cite[Sec.~0.7.3]{matrixanalysis}, \[
\scriptsize \begin{bmatrix} \bA & \bB \\ \bC & \bD \end{bmatrix}^{-1}
=
\begin{bmatrix} 
  (\bA - \bB\bD^{-1}\bC)^{-1} & \bA^{-1}\bB(\bC\bA^{-1}\bB-\bD)^{-1}
  \\ (\bC\bA^{-1}\bB-\bD)^{-1} \bC \bA^{-1} & (\bD -
  \bC\bA^{-1}\bB)^{-1}
\end{bmatrix}.\]}

We have already assumed that the edges of the graph are ordered such
that
\begin{align}
\bB = \begin{bmatrix} \bB''' \\ \bBom \end{bmatrix},
\end{align}
for some $\bB'''$. Define
\begin{align}\label{eq:bdeldef}
  \bDel & = \bL - \bLom = \bB^T\bQ^{-1}\bB  - \bB_\Omega^T \bQ_\Omega^{-1} \bB_\Omega \nonumber \\
  & = \bB^T\bQ^{-1}\bB  - \bB^T \begin{bmatrix}   \bzero \\ \bI \end{bmatrix} \bQ_\Omega^{-1} \begin{bmatrix}  \bzero & \bI \end{bmatrix} \bB  \nonumber \\ & =\bB^T \pp{ \bQ^{-1}   -   \begin{bmatrix}   \bzero \\ \bI \end{bmatrix} \bQ_\Omega^{-1} \begin{bmatrix}   \bzero & \bI \end{bmatrix} }\bB  .
\end{align}
By recalling that $\bQ_\Omega$ is the lower-right $N_\Omega\times
N_\Omega$ submatrix of $\bQ$, rewriting $\bQ^{-1}$ using a Schur
complement, and performing a calculation similar to that in
(\ref{eq:LppmL}), we conclude that $\bDel\succcurlyeq\bzero$.  Also,
because of (\ref{eq:Lomdecomp}),
\begin{align}
  \bLom \bEbom = \bzero.
\end{align}
Using (\ref{eq:defKb}) and (\ref{eq:Ka2}), the difference between the
inverse kernels can then be written,
\begin{align}
  & \bK_a^{-1}-\bK_b^{-1} \nonumber \\
  & =  \bZom^T\bL\bZom - \bZom^T\bL\bEbom (\bEbom^T \bL \bEbom)^{-1} \bEbom^T \bL \bZom \nonumber \\ & \qquad - \bZom^T \bLom \bZom \nonumber \\
  & =
  \bZom^T \pp{ \bL  -  \bLom -  \bL\bEbom (\bEbom^T \bL \bEbom)^{-1} \bEbom^T \bL   } \bZom \nonumber \\
  & = 
  \bZom^T \Big( \bDel -  (\bLom+\bDel) \bEbom (\bEbom^T (\bLom+\bDel) \bEbom)^{-1}  \nonumber \\ & \qquad \cdot \bEbom^T (\bLom+\bDel)   \Big) \bZom \nonumber \\
  & = 
  \bZom^T \pp{ \bDel -  \bDel  \bEbom (\bEbom^T  \bDel \bEbom)^{-1} \bEbom^T \bDel   } \bZom \nonumber \\
  & = 
  \bZom^T \bDel^{1/2} \pp{ \bI -  \bDel^{1/2}  \bEbom (\bEbom^T  \bDel \bEbom)^{-1} \bEbom^T \bDel^{1/2}   } \nonumber \\ & \qquad \cdot \bDel^{1/2}  \bZom. \label{eq:Kdiff}
\end{align}
The matrix inside the parenthesis after the last equality in
(\ref{eq:Kdiff}) is the orthogonal projection onto the orthogonal
complement of the column space of $ \bDel^{1/2} \bEbom$. Therefore,
(\ref{eq:Kdiff}) is positive semidefinite, and the desired result
follows: the beamforming accuracy in case (a) is always better than,
or equal to, that in case (b).

As a final remark, we comment on the implications of monotonicity
(Section~\ref{sec:mono}) on the beamforming gain, when adding more
measurements.  Consider case (a). Let $V_a$ be the nominal variance,
and $V''_a$ be the variance after the addition of supplementary
measurements.  From (\ref{eq:monotonicity}) and (\ref{eq:wcvara}) it
is then immediate that $V''_a\le V_a$, for any permissible $\{a_n\}$.
This effect must not, of course, be conflated with the conclusion from
the comparison between cases (a) and (b) derived above.

\section{Examples}

The estimation errors in $\{\hat\phi_n\}$ can grow fast when scaling
up the network. Yet, the beamforming accuracy, for any subset
$\Omega$, is always better when all measurements are used.  This is
best illustrated through the study of some special cases.

In all examples, we assume that the measurement noises are
uncorrelated, and set the noise variance to one: $\bQ=\bI$. (In the
numerical illustrations, we scale $\bQ$.)  In this case, $\bL$ is the
standard Laplacian. When reading the examples, it is useful to keep in
mind its equivalent definition: ${\bL=\bD-\bA}$, where $\bA$ is the
graph adjacency matrix (${[\bA]_{nn'}=1}$ if $n$ and $n'$ are
connected, and zero otherwise) and $\bD$ is a diagonal matrix with
$\bA\bu$ on its diagonal.

\subsection{Linear (Radio Stripe) Topology}\label{sec:stripe}

First we consider the line (radio stripe) topology in
Figure~\ref{fig:lrs}, where measurements are only conducted between
neighboring antennas.  The Laplacian is immediate from its definition:
\begin{align}\label{eq:ringL}
  \bL = \begin{bmatrix}
    1 & -1 & 0 & \cdots & \cdots & \cdots & \cdots & 0\\
    -1 & 2 & -1 & 0 & & & & \vdots\\
    0 & -1 & 2 & -1 & \ddots & & & \vdots\\
    \vdots & 0 & \ddots & \ddots & \ddots & \ddots & & \vdots\\
    \vdots & & \ddots & \ddots & \ddots & \ddots & 0 & \vdots\\
    \vdots & & & \ddots & -1 & 2 & -1 & 0\\
    \vdots & & & & 0 & -1 & 2 & -1\\
    0 & \cdots & \cdots  & \cdots & \cdots & 0 & -1 & 1\\
  \end{bmatrix}.
\end{align}
Let 
\begin{align}\label{eq:lineeigv}
  \by_n = 
  \begin{bmatrix}  
    \cos\pp{  \frac{1}{2} \frac{(n-1)\pi}{N}    }  \\ 
    \cos\pp{  \frac{3}{2} \frac{(n-1)\pi}{N}    }   \\ 
    \vdots \\ 
    \cos\pp{  \frac{2N-3}{2} \frac{(n-1)\pi}{N}    }   \\ 
    \cos\pp{  \frac{2N-1}{2} \frac{(n-1)\pi}{N}    }   
  \end{bmatrix},
\end{align}
for $n=1,...,N$. A direct but  tedious calculation, or the use
of results from \cite{cvetkovic1980spectra}, shows that $\{ \by_n \}$
are mutually orthogonal and that they are eigenvectors of $\bL$:
\begin{align}
  \bL \by_1 & = \bzero  \\
  \bL\by_{n} & =  [\bLam]_{(n-1)(n-1)} \by_{n},\qquad n=2,...,N,  \\
  \by_n^T\by_{n'} & =0,\qquad n\neq n',
\end{align}
where the corresponding eigenvalues are
\begin{align}
  [\bLam]_{(n-1)(n-1)} = 4\sin^2\pp{\frac{(n-1)\pi}{2N}}.
\end{align}
Note that the first eigenvalue is zero; actually, $\by_1=\bu$.  Taken
together, this means that $\bZ$ can be written as $\bZ=[\bz_2 \ \cdots
  \ \bz_N]$ where
\begin{align}
  \bz_n  & = \frac{\by_n}{\Vert\by_n\Vert}    .
\end{align}
In particular, (\ref{eq:ZLZL}) holds.
 
It follows that,
\begin{align}\label{eq:zl}
  \cov{\hat\bphi} & = \bZ(\bZ^T\bL\bZ)^{-1}\bZ^T \nonumber \\
  & = \sum_{n=2}^N \frac{\bz_n\bz_n^T}{[\bLam]_{(n-1)(n-1)} } =
  \frac{1}{4}  \sum_{n=2}^N \frac{\by_n\by_n^T}{\sin^2\pp{\frac{(n-1)\pi}{2N}} \Vert\by_n\Vert^2} .
\end{align}
Now use (\ref{eq:lineeigv}) to write the $n$th diagonal element of
$\cov{\hat\bphi}$ as
\begin{align}\label{eq:varphi1}
  \var{\hat\phi_n} & = \frac{1}{4} \sum_{n'=2}^N \frac{ \cos^2\pp{
      \pp{2n-1} \frac{(n'-1)\pi}{2N} }  }{\sin^2\pp{\frac{(n'-1)\pi}{2N}} \Vert\by_{n'}\Vert^2} .
\end{align}
The right hand side of (\ref{eq:varphi1}) can be lower-bounded by
retaining only the first two terms, for which $n'=2$ and
$n'=3$. (Keeping only the first term turns out to be insufficient.)
The sum of the numerators corresponding to $n'=2$ and $n'=3$ is
lower-bounded, uniformly over $n$, as follows:
\begin{align}
  & \quad \cos^2\pp{ (2n-1)\frac{\pi}{2N}}  + \cos^2\pp{ (2n-1)\frac{\pi}{N}} \nonumber \\
  & = \frac{1}{2} \ppb{ 1 + \cos\pp{(2n-1)\frac{\pi}{N}} + 2\cos^2\pp{ (2n-1)\frac{\pi}{N}}} \nonumber \\
  & = \frac{1}{2} \ppb{1 + \frac{1}{2}\cos\pp{(2n-1)\frac{\pi}{N}}}^2 + \frac{7}{8} \cos^2\pp{(2n-1)\frac{\pi}{N}} \nonumber \\
  & \ge \frac{1}{8}.
\end{align}
From (\ref{eq:lineeigv}) it is immediate that   
\begin{align}
  \Vert\by_n\Vert^2 & \le N, \qquad n=1,...,N.
\end{align}
Since $\sin(x) \le x$ for $ x\ge 0$, the denominator of the terms
inside the sum of (\ref{eq:varphi1}) for $n'=2$ and $n'=3$ is
upper-bounded by,
\begin{align}
  \sin^2\pp{\frac{(n'-1)\pi}{2N} } \Vert\by_{n'}\Vert^2 \le   \pp{\frac{\pi}{N}}^2 N = \frac{\pi^2}{N}.
\end{align}
Putting these bounds together we find that uniformly over $n$,
\begin{align} \label{eq:varbound}
  \var{\hat\phi_n} \ge   \frac{N}{32\pi^2}.
\end{align}
This shows that the smallest value among $\var{\hat\phi_n}\to \infty$
as $N\to\infty$ (rather quickly).  Despite this, we know, from the
analysis in Section~\ref{sec:gaincomp}, that the beamforming accuracy
for any fixed subset $\Omega$ cannot decrease when including more
antennas in the calibration process.  In this particular example with
a line topology, it turns out that $\bK_a=\bK_b$ (see below), but we
will see an example later (Section~\ref{sec:lis2d}) where $\bK_b$ is
strictly larger than $\bK_a$.

To see why $\bK_a=\bK_b$ in the present example, consider first the
case that $\Omega=\{1,...,N_\Omega \}$, and look at the penultimate
line of (\ref{eq:Kdiff}). Let $\bPsi=\bDel - \bDel \bEbom (\bEbom^T
\bDel \bEbom)^{-1} \bEbom^T \bDel$ and note that $\bPsi
\bEbom=\bzero$; likewise, $\bPsi\bu =\bzero$.  Also note that $\bDel$
has zeros in its $(N_\Omega-1) \times (N_\Omega-1)$ upper-left corner;
so has $\bPsi$.  Since $\bu$ is linearly independent of the columns of
$\bEbom$, the only possibility is $\bPsi=\bzero$, which, by
(\ref{eq:Kdiff}), implies $\bK_a=\bK_b$.  The case when $\Omega$
consists of a different set of consecutive indices can be handled
similarly.

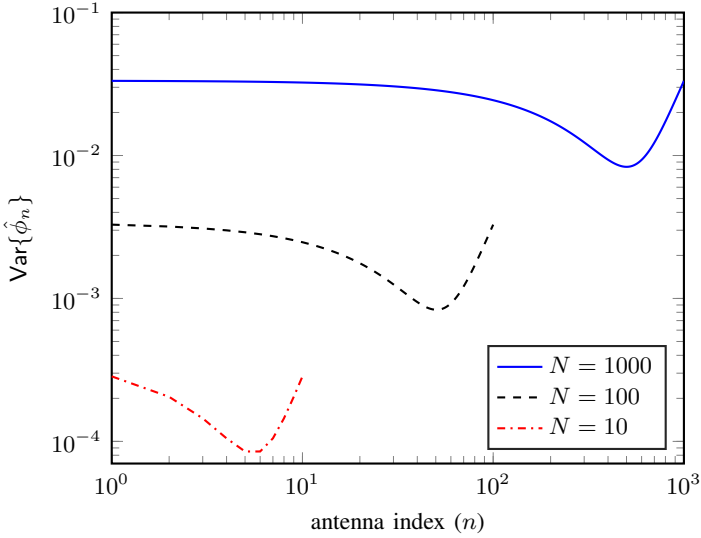
\begin{figure}[t!]
  \centerline{\scalebox{1}{
\begin{tikzpicture}

\begin{axis}[%
thick,
width=7.607cm,
height=6cm,
at={(0cm,0cm)},
scale only axis,
xmode=log,
xmin=1,
xmax=1000,
xminorticks=true,
xlabel style={font=\color{white!15!black}},
xlabel={antenna index ($n$)},
ymode=log,
ymin=7e-05,
ymax=0.1,
yminorticks=true,
ylabel style={font=\color{white!15!black}},
ylabel={$\var{\hat\phi_n}$},
axis background/.style={fill=white},
legend style={at={(0.97,0.03)}, anchor=south east, legend cell align=left, align=left, draw=white!15!black},
ylabel near ticks,
xlabel near ticks,
ylabel style={font=\small},
xlabel style={font=\small},
legend style={font=\small},,
ticklabel style={font=\small}
]
\addplot [color=blue]
  table[row sep=crcr]{%
1	0.0332833499993822\\
2	0.0331835499993822\\
3	0.0330839499993823\\
4	0.0329845499993823\\
5	0.0328853499993823\\
6	0.0327863499993823\\
7	0.0326875499993824\\
8	0.0325889499993825\\
9	0.0324905499993825\\
10	0.0323923499993826\\
11	0.0322943499993827\\
12	0.0321965499993827\\
13	0.0320989499993828\\
14	0.0320015499993829\\
15	0.031904349999383\\
16	0.0318073499993832\\
17	0.0317105499993833\\
18	0.0316139499993834\\
19	0.0315175499993835\\
20	0.0314213499993837\\
21	0.0313253499993838\\
22	0.031229549999384\\
23	0.0311339499993842\\
24	0.0310385499993844\\
25	0.0309433499993845\\
26	0.0308483499993847\\
27	0.0307535499993849\\
28	0.0306589499993851\\
29	0.0305645499993854\\
30	0.0304703499993856\\
31	0.0303763499993858\\
32	0.0302825499993861\\
33	0.0301889499993863\\
34	0.0300955499993866\\
35	0.0300023499993868\\
36	0.0299093499993871\\
37	0.0298165499993874\\
38	0.0297239499993877\\
39	0.029631549999388\\
40	0.0295393499993882\\
41	0.0294473499993886\\
42	0.0293555499993889\\
43	0.0292639499993892\\
44	0.0291725499993895\\
45	0.0290813499993899\\
46	0.0289903499993902\\
47	0.0288995499993905\\
48	0.0288089499993909\\
49	0.0287185499993913\\
50	0.0286283499993916\\
51	0.028538349999392\\
52	0.0284485499993924\\
53	0.0283589499993928\\
54	0.0282695499993932\\
55	0.0281803499993936\\
56	0.028091349999394\\
57	0.0280025499993944\\
58	0.0279139499993949\\
59	0.0278255499993953\\
60	0.0277373499993957\\
61	0.0276493499993962\\
62	0.0275615499993966\\
63	0.0274739499993971\\
64	0.0273865499993976\\
65	0.0272993499993981\\
66	0.0272123499993985\\
67	0.027125549999399\\
68	0.0270389499993995\\
69	0.0269525499994\\
70	0.0268663499994006\\
71	0.0267803499994011\\
72	0.0266945499994016\\
73	0.0266089499994021\\
74	0.0265235499994027\\
75	0.0264383499994032\\
76	0.0263533499994038\\
77	0.0262685499994043\\
78	0.0261839499994049\\
79	0.0260995499994055\\
80	0.0260153499994061\\
81	0.0259313499994067\\
82	0.0258475499994073\\
83	0.0257639499994079\\
84	0.0256805499994085\\
85	0.0255973499994091\\
86	0.0255143499994097\\
87	0.0254315499994103\\
88	0.025348949999411\\
89	0.0252665499994116\\
90	0.0251843499994123\\
91	0.0251023499994129\\
92	0.0250205499994136\\
93	0.0249389499994143\\
94	0.024857549999415\\
95	0.0247763499994156\\
96	0.0246953499994163\\
97	0.024614549999417\\
98	0.0245339499994177\\
99	0.0244535499994184\\
100	0.0243733499994192\\
101	0.0242933499994199\\
102	0.0242135499994206\\
103	0.0241339499994214\\
104	0.0240545499994221\\
105	0.0239753499994229\\
106	0.0238963499994236\\
107	0.0238175499994244\\
108	0.0237389499994252\\
109	0.0236605499994259\\
110	0.0235823499994267\\
111	0.0235043499994275\\
112	0.0234265499994283\\
113	0.0233489499994291\\
114	0.0232715499994299\\
115	0.0231943499994307\\
116	0.0231173499994316\\
117	0.0230405499994324\\
118	0.0229639499994332\\
119	0.0228875499994341\\
120	0.0228113499994349\\
121	0.0227353499994358\\
122	0.0226595499994366\\
123	0.0225839499994375\\
124	0.0225085499994384\\
125	0.0224333499994393\\
126	0.0223583499994401\\
127	0.022283549999441\\
128	0.0222089499994419\\
129	0.0221345499994428\\
130	0.0220603499994437\\
131	0.0219863499994447\\
132	0.0219125499994456\\
133	0.0218389499994465\\
134	0.0217655499994475\\
135	0.0216923499994484\\
136	0.0216193499994493\\
137	0.0215465499994503\\
138	0.0214739499994513\\
139	0.0214015499994522\\
140	0.0213293499994532\\
141	0.0212573499994542\\
142	0.0211855499994552\\
143	0.0211139499994562\\
144	0.0210425499994572\\
145	0.0209713499994582\\
146	0.0209003499994592\\
147	0.0208295499994602\\
148	0.0207589499994612\\
149	0.0206885499994622\\
150	0.0206183499994632\\
151	0.0205483499994643\\
152	0.0204785499994653\\
153	0.0204089499994664\\
154	0.0203395499994674\\
155	0.0202703499994685\\
156	0.0202013499994696\\
157	0.0201325499994706\\
158	0.0200639499994717\\
159	0.0199955499994728\\
160	0.0199273499994739\\
161	0.019859349999475\\
162	0.0197915499994761\\
163	0.0197239499994772\\
164	0.0196565499994783\\
165	0.0195893499994794\\
166	0.0195223499994805\\
167	0.0194555499994816\\
168	0.0193889499994827\\
169	0.0193225499994839\\
170	0.019256349999485\\
171	0.0191903499994862\\
172	0.0191245499994873\\
173	0.0190589499994885\\
174	0.0189935499994896\\
175	0.0189283499994908\\
176	0.018863349999492\\
177	0.0187985499994932\\
178	0.0187339499994943\\
179	0.0186695499994955\\
180	0.0186053499994967\\
181	0.0185413499994979\\
182	0.0184775499994991\\
183	0.0184139499995003\\
184	0.0183505499995015\\
185	0.0182873499995028\\
186	0.018224349999504\\
187	0.0181615499995052\\
188	0.0180989499995064\\
189	0.0180365499995077\\
190	0.0179743499995089\\
191	0.0179123499995102\\
192	0.0178505499995114\\
193	0.0177889499995127\\
194	0.017727549999514\\
195	0.0176663499995152\\
196	0.0176053499995165\\
197	0.0175445499995178\\
198	0.017483949999519\\
199	0.0174235499995203\\
200	0.0173633499995216\\
201	0.0173033499995229\\
202	0.0172435499995242\\
203	0.0171839499995255\\
204	0.0171245499995268\\
205	0.0170653499995281\\
206	0.0170063499995295\\
207	0.0169475499995308\\
208	0.0168889499995321\\
209	0.0168305499995334\\
210	0.0167723499995348\\
211	0.0167143499995361\\
212	0.0166565499995374\\
213	0.0165989499995388\\
214	0.0165415499995401\\
215	0.0164843499995415\\
216	0.0164273499995429\\
217	0.0163705499995442\\
218	0.0163139499995456\\
219	0.016257549999547\\
220	0.0162013499995483\\
221	0.0161453499995497\\
222	0.0160895499995511\\
223	0.0160339499995525\\
224	0.0159785499995539\\
225	0.0159233499995553\\
226	0.0158683499995567\\
227	0.0158135499995581\\
228	0.0157589499995595\\
229	0.0157045499995609\\
230	0.0156503499995623\\
231	0.0155963499995637\\
232	0.0155425499995651\\
233	0.0154889499995666\\
234	0.015435549999568\\
235	0.0153823499995694\\
236	0.0153293499995709\\
237	0.0152765499995723\\
238	0.0152239499995737\\
239	0.0151715499995752\\
240	0.0151193499995766\\
241	0.0150673499995781\\
242	0.0150155499995796\\
243	0.014963949999581\\
244	0.0149125499995825\\
245	0.014861349999584\\
246	0.0148103499995854\\
247	0.0147595499995869\\
248	0.0147089499995884\\
249	0.0146585499995899\\
250	0.0146083499995914\\
251	0.0145583499995929\\
252	0.0145085499995944\\
253	0.0144589499995958\\
254	0.0144095499995973\\
255	0.0143603499995989\\
256	0.0143113499996004\\
257	0.0142625499996019\\
258	0.0142139499996034\\
259	0.0141655499996049\\
260	0.0141173499996064\\
261	0.0140693499996079\\
262	0.0140215499996095\\
263	0.013973949999611\\
264	0.0139265499996125\\
265	0.0138793499996141\\
266	0.0138323499996156\\
267	0.0137855499996171\\
268	0.0137389499996187\\
269	0.0136925499996202\\
270	0.0136463499996218\\
271	0.0136003499996234\\
272	0.0135545499996249\\
273	0.0135089499996265\\
274	0.013463549999628\\
275	0.0134183499996296\\
276	0.0133733499996312\\
277	0.0133285499996328\\
278	0.0132839499996343\\
279	0.0132395499996359\\
280	0.0131953499996375\\
281	0.0131513499996391\\
282	0.0131075499996407\\
283	0.0130639499996423\\
284	0.0130205499996438\\
285	0.0129773499996454\\
286	0.012934349999647\\
287	0.0128915499996486\\
288	0.0128489499996502\\
289	0.0128065499996519\\
290	0.0127643499996535\\
291	0.0127223499996551\\
292	0.0126805499996567\\
293	0.0126389499996583\\
294	0.0125975499996599\\
295	0.0125563499996616\\
296	0.0125153499996632\\
297	0.0124745499996648\\
298	0.0124339499996664\\
299	0.0123935499996681\\
300	0.0123533499996697\\
301	0.0123133499996714\\
302	0.012273549999673\\
303	0.0122339499996746\\
304	0.0121945499996763\\
305	0.0121553499996779\\
306	0.0121163499996796\\
307	0.0120775499996812\\
308	0.0120389499996829\\
309	0.0120005499996846\\
310	0.0119623499996862\\
311	0.0119243499996879\\
312	0.0118865499996896\\
313	0.0118489499996912\\
314	0.0118115499996929\\
315	0.0117743499996946\\
316	0.0117373499996962\\
317	0.0117005499996979\\
318	0.0116639499996996\\
319	0.0116275499997013\\
320	0.011591349999703\\
321	0.0115553499997047\\
322	0.0115195499997063\\
323	0.011483949999708\\
324	0.0114485499997097\\
325	0.0114133499997114\\
326	0.0113783499997131\\
327	0.0113435499997148\\
328	0.0113089499997165\\
329	0.0112745499997182\\
330	0.0112403499997199\\
331	0.0112063499997217\\
332	0.0111725499997234\\
333	0.0111389499997251\\
334	0.0111055499997268\\
335	0.0110723499997285\\
336	0.0110393499997302\\
337	0.011006549999732\\
338	0.0109739499997337\\
339	0.0109415499997354\\
340	0.0109093499997371\\
341	0.0108773499997389\\
342	0.0108455499997406\\
343	0.0108139499997423\\
344	0.0107825499997441\\
345	0.0107513499997458\\
346	0.0107203499997476\\
347	0.0106895499997493\\
348	0.010658949999751\\
349	0.0106285499997528\\
350	0.0105983499997545\\
351	0.0105683499997563\\
352	0.010538549999758\\
353	0.0105089499997598\\
354	0.0104795499997616\\
355	0.0104503499997633\\
356	0.0104213499997651\\
357	0.0103925499997668\\
358	0.0103639499997686\\
359	0.0103355499997704\\
360	0.0103073499997721\\
361	0.0102793499997739\\
362	0.0102515499997757\\
363	0.0102239499997774\\
364	0.0101965499997792\\
365	0.010169349999781\\
366	0.0101423499997828\\
367	0.0101155499997846\\
368	0.0100889499997863\\
369	0.0100625499997881\\
370	0.0100363499997899\\
371	0.0100103499997917\\
372	0.00998454999979347\\
373	0.00995894999979525\\
374	0.00993354999979704\\
375	0.00990834999979883\\
376	0.00988334999980063\\
377	0.00985854999980242\\
378	0.00983394999980421\\
379	0.00980954999980601\\
380	0.0097853499998078\\
381	0.0097613499998096\\
382	0.0097375499998114\\
383	0.0097139499998132\\
384	0.009690549999815\\
385	0.0096673499998168\\
386	0.00964434999981861\\
387	0.00962154999982041\\
388	0.00959894999982222\\
389	0.00957654999982402\\
390	0.00955434999982583\\
391	0.00953234999982764\\
392	0.00951054999982945\\
393	0.00948894999983126\\
394	0.00946754999983307\\
395	0.00944634999983489\\
396	0.0094253499998367\\
397	0.00940454999983852\\
398	0.00938394999984033\\
399	0.00936354999984215\\
400	0.00934334999984397\\
401	0.00932334999984579\\
402	0.00930354999984761\\
403	0.00928394999984943\\
404	0.00926454999985125\\
405	0.00924534999985307\\
406	0.0092263499998549\\
407	0.00920754999985672\\
408	0.00918894999985854\\
409	0.00917054999986037\\
410	0.00915234999986219\\
411	0.00913434999986402\\
412	0.00911654999986584\\
413	0.00909894999986767\\
414	0.0090815499998695\\
415	0.00906434999987132\\
416	0.00904734999987315\\
417	0.00903054999987498\\
418	0.00901394999987681\\
419	0.00899754999987864\\
420	0.00898134999988046\\
421	0.00896534999988229\\
422	0.00894954999988412\\
423	0.00893394999988596\\
424	0.00891854999988779\\
425	0.00890334999988962\\
426	0.00888834999989145\\
427	0.00887354999989328\\
428	0.00885894999989511\\
429	0.00884454999989694\\
430	0.00883034999989878\\
431	0.00881634999990061\\
432	0.00880254999990245\\
433	0.00878894999990428\\
434	0.00877554999990612\\
435	0.00876234999990795\\
436	0.00874934999990978\\
437	0.00873654999991162\\
438	0.00872394999991345\\
439	0.00871154999991528\\
440	0.00869934999991712\\
441	0.00868734999991895\\
442	0.00867554999992079\\
443	0.00866394999992262\\
444	0.00865254999992446\\
445	0.00864134999992629\\
446	0.00863034999992813\\
447	0.00861954999992996\\
448	0.00860894999993179\\
449	0.00859854999993362\\
450	0.00858834999993546\\
451	0.00857834999993729\\
452	0.00856854999993912\\
453	0.00855894999994096\\
454	0.00854954999994279\\
455	0.00854034999994462\\
456	0.00853134999994645\\
457	0.00852254999994829\\
458	0.00851394999995012\\
459	0.00850554999995195\\
460	0.00849734999995379\\
461	0.00848934999995562\\
462	0.00848154999995745\\
463	0.00847394999995928\\
464	0.00846654999996111\\
465	0.00845934999996294\\
466	0.00845234999996477\\
467	0.00844554999996659\\
468	0.00843894999996842\\
469	0.00843254999997024\\
470	0.00842634999997207\\
471	0.00842034999997389\\
472	0.00841454999997572\\
473	0.00840894999997754\\
474	0.00840354999997936\\
475	0.00839834999998119\\
476	0.00839334999998301\\
477	0.00838854999998483\\
478	0.00838394999998665\\
479	0.00837954999998846\\
480	0.00837534999999028\\
481	0.0083713499999921\\
482	0.00836754999999392\\
483	0.00836394999999573\\
484	0.00836054999999754\\
485	0.00835734999999935\\
486	0.00835435000000117\\
487	0.00835155000000298\\
488	0.00834895000000479\\
489	0.0083465500000066\\
490	0.00834435000000841\\
491	0.00834235000001022\\
492	0.00834055000001202\\
493	0.00833895000001383\\
494	0.00833755000001563\\
495	0.00833635000001744\\
496	0.00833535000001924\\
497	0.00833455000002104\\
498	0.00833395000002284\\
499	0.00833355000002464\\
500	0.00833335000002644\\
501	0.00833335000002823\\
502	0.00833355000003003\\
503	0.00833395000003182\\
504	0.00833455000003362\\
505	0.00833535000003541\\
506	0.0083363500000372\\
507	0.008337550000039\\
508	0.00833895000004078\\
509	0.00834055000004257\\
510	0.00834235000004436\\
511	0.00834435000004615\\
512	0.00834655000004793\\
513	0.00834895000004971\\
514	0.0083515500000515\\
515	0.00835435000005328\\
516	0.00835735000005506\\
517	0.00836055000005684\\
518	0.00836395000005861\\
519	0.00836755000006039\\
520	0.00837135000006217\\
521	0.00837535000006394\\
522	0.00837955000006571\\
523	0.00838395000006748\\
524	0.00838855000006926\\
525	0.00839335000007102\\
526	0.00839835000007279\\
527	0.00840355000007456\\
528	0.00840895000007632\\
529	0.00841455000007808\\
530	0.00842035000007985\\
531	0.0084263500000816\\
532	0.00843255000008336\\
533	0.00843895000008512\\
534	0.00844555000008687\\
535	0.00845235000008863\\
536	0.00845935000009038\\
537	0.00846655000009213\\
538	0.00847395000009388\\
539	0.00848155000009563\\
540	0.00848935000009737\\
541	0.00849735000009912\\
542	0.00850555000010086\\
543	0.0085139500001026\\
544	0.00852255000010434\\
545	0.00853135000010608\\
546	0.00854035000010782\\
547	0.00854955000010955\\
548	0.00855895000011128\\
549	0.00856855000011302\\
550	0.00857835000011475\\
551	0.00858835000011647\\
552	0.0085985500001182\\
553	0.00860895000011992\\
554	0.00861955000012165\\
555	0.00863035000012337\\
556	0.00864135000012508\\
557	0.0086525500001268\\
558	0.00866395000012852\\
559	0.00867555000013023\\
560	0.00868735000013194\\
561	0.00869935000013365\\
562	0.00871155000013536\\
563	0.00872395000013707\\
564	0.00873655000013877\\
565	0.00874935000014048\\
566	0.00876235000014218\\
567	0.00877555000014388\\
568	0.00878895000014557\\
569	0.00880255000014727\\
570	0.00881635000014896\\
571	0.00883035000015065\\
572	0.00884455000015234\\
573	0.00885895000015403\\
574	0.00887355000015571\\
575	0.00888835000015739\\
576	0.00890335000015907\\
577	0.00891855000016075\\
578	0.00893395000016243\\
579	0.0089495500001641\\
580	0.00896535000016577\\
581	0.00898135000016744\\
582	0.00899755000016911\\
583	0.00901395000017077\\
584	0.00903055000017244\\
585	0.0090473500001741\\
586	0.00906435000017576\\
587	0.00908155000017741\\
588	0.00909895000017906\\
589	0.00911655000018071\\
590	0.00913435000018236\\
591	0.00915235000018401\\
592	0.00917055000018566\\
593	0.0091889500001873\\
594	0.00920755000018894\\
595	0.00922635000019057\\
596	0.00924535000019221\\
597	0.00926455000019385\\
598	0.00928395000019548\\
599	0.00930355000019711\\
600	0.00932335000019873\\
601	0.00934335000020036\\
602	0.00936355000020198\\
603	0.0093839500002036\\
604	0.00940455000020522\\
605	0.00942535000020683\\
606	0.00944635000020845\\
607	0.00946755000021006\\
608	0.00948895000021167\\
609	0.00951055000021327\\
610	0.00953235000021488\\
611	0.00955435000021648\\
612	0.00957655000021808\\
613	0.00959895000021968\\
614	0.00962155000022127\\
615	0.00964435000022286\\
616	0.00966735000022445\\
617	0.00969055000022604\\
618	0.00971395000022762\\
619	0.0097375500002292\\
620	0.00976135000023078\\
621	0.00978535000023236\\
622	0.00980955000023393\\
623	0.0098339500002355\\
624	0.00985855000023707\\
625	0.00988335000023864\\
626	0.0099083500002402\\
627	0.00993355000024176\\
628	0.00995895000024332\\
629	0.00998455000024487\\
630	0.0100103500002464\\
631	0.010036350000248\\
632	0.0100625500002495\\
633	0.0100889500002511\\
634	0.0101155500002526\\
635	0.0101423500002541\\
636	0.0101693500002557\\
637	0.0101965500002572\\
638	0.0102239500002587\\
639	0.0102515500002603\\
640	0.0102793500002618\\
641	0.0103073500002633\\
642	0.0103355500002648\\
643	0.0103639500002663\\
644	0.0103925500002679\\
645	0.0104213500002694\\
646	0.0104503500002709\\
647	0.0104795500002724\\
648	0.0105089500002739\\
649	0.0105385500002754\\
650	0.0105683500002769\\
651	0.0105983500002784\\
652	0.0106285500002799\\
653	0.0106589500002814\\
654	0.0106895500002828\\
655	0.0107203500002843\\
656	0.0107513500002858\\
657	0.0107825500002873\\
658	0.0108139500002888\\
659	0.0108455500002902\\
660	0.0108773500002917\\
661	0.0109093500002932\\
662	0.0109415500002946\\
663	0.0109739500002961\\
664	0.0110065500002975\\
665	0.011039350000299\\
666	0.0110723500003004\\
667	0.0111055500003019\\
668	0.0111389500003033\\
669	0.0111725500003048\\
670	0.0112063500003062\\
671	0.0112403500003076\\
672	0.0112745500003091\\
673	0.0113089500003105\\
674	0.0113435500003119\\
675	0.0113783500003133\\
676	0.0114133500003148\\
677	0.0114485500003162\\
678	0.0114839500003176\\
679	0.011519550000319\\
680	0.0115553500003204\\
681	0.0115913500003218\\
682	0.0116275500003232\\
683	0.0116639500003246\\
684	0.011700550000326\\
685	0.0117373500003274\\
686	0.0117743500003288\\
687	0.0118115500003301\\
688	0.0118489500003315\\
689	0.0118865500003329\\
690	0.0119243500003343\\
691	0.0119623500003356\\
692	0.012000550000337\\
693	0.0120389500003384\\
694	0.0120775500003397\\
695	0.0121163500003411\\
696	0.0121553500003424\\
697	0.0121945500003438\\
698	0.0122339500003451\\
699	0.0122735500003464\\
700	0.0123133500003478\\
701	0.0123533500003491\\
702	0.0123935500003504\\
703	0.0124339500003518\\
704	0.0124745500003531\\
705	0.0125153500003544\\
706	0.0125563500003557\\
707	0.012597550000357\\
708	0.0126389500003583\\
709	0.0126805500003596\\
710	0.0127223500003609\\
711	0.0127643500003622\\
712	0.0128065500003635\\
713	0.0128489500003648\\
714	0.0128915500003661\\
715	0.0129343500003674\\
716	0.0129773500003687\\
717	0.01302055000037\\
718	0.0130639500003712\\
719	0.0131075500003725\\
720	0.0131513500003738\\
721	0.013195350000375\\
722	0.0132395500003763\\
723	0.0132839500003775\\
724	0.0133285500003788\\
725	0.01337335000038\\
726	0.0134183500003813\\
727	0.0134635500003825\\
728	0.0135089500003838\\
729	0.013554550000385\\
730	0.0136003500003862\\
731	0.0136463500003874\\
732	0.0136925500003887\\
733	0.0137389500003899\\
734	0.0137855500003911\\
735	0.0138323500003923\\
736	0.0138793500003935\\
737	0.0139265500003947\\
738	0.0139739500003959\\
739	0.0140215500003971\\
740	0.0140693500003983\\
741	0.0141173500003995\\
742	0.0141655500004007\\
743	0.0142139500004019\\
744	0.014262550000403\\
745	0.0143113500004042\\
746	0.0143603500004054\\
747	0.0144095500004065\\
748	0.0144589500004077\\
749	0.0145085500004089\\
750	0.01455835000041\\
751	0.0146083500004112\\
752	0.0146585500004123\\
753	0.0147089500004134\\
754	0.0147595500004146\\
755	0.0148103500004157\\
756	0.0148613500004168\\
757	0.014912550000418\\
758	0.0149639500004191\\
759	0.0150155500004202\\
760	0.0150673500004213\\
761	0.0151193500004224\\
762	0.0151715500004235\\
763	0.0152239500004246\\
764	0.0152765500004257\\
765	0.0153293500004268\\
766	0.0153823500004279\\
767	0.015435550000429\\
768	0.0154889500004301\\
769	0.0155425500004311\\
770	0.0155963500004322\\
771	0.0156503500004333\\
772	0.0157045500004343\\
773	0.0157589500004354\\
774	0.0158135500004364\\
775	0.0158683500004375\\
776	0.0159233500004385\\
777	0.0159785500004396\\
778	0.0160339500004406\\
779	0.0160895500004417\\
780	0.0161453500004427\\
781	0.0162013500004437\\
782	0.0162575500004447\\
783	0.0163139500004458\\
784	0.0163705500004468\\
785	0.0164273500004478\\
786	0.0164843500004488\\
787	0.0165415500004498\\
788	0.0165989500004508\\
789	0.0166565500004518\\
790	0.0167143500004528\\
791	0.0167723500004537\\
792	0.0168305500004547\\
793	0.0168889500004557\\
794	0.0169475500004567\\
795	0.0170063500004576\\
796	0.0170653500004586\\
797	0.0171245500004596\\
798	0.0171839500004605\\
799	0.0172435500004615\\
800	0.0173033500004624\\
801	0.0173633500004634\\
802	0.0174235500004643\\
803	0.0174839500004652\\
804	0.0175445500004661\\
805	0.0176053500004671\\
806	0.017666350000468\\
807	0.0177275500004689\\
808	0.0177889500004698\\
809	0.0178505500004707\\
810	0.0179123500004716\\
811	0.0179743500004725\\
812	0.0180365500004734\\
813	0.0180989500004743\\
814	0.0181615500004752\\
815	0.018224350000476\\
816	0.0182873500004769\\
817	0.0183505500004778\\
818	0.0184139500004786\\
819	0.0184775500004795\\
820	0.0185413500004804\\
821	0.0186053500004812\\
822	0.0186695500004821\\
823	0.0187339500004829\\
824	0.0187985500004837\\
825	0.0188633500004846\\
826	0.0189283500004854\\
827	0.0189935500004862\\
828	0.019058950000487\\
829	0.0191245500004878\\
830	0.0191903500004886\\
831	0.0192563500004894\\
832	0.0193225500004902\\
833	0.019388950000491\\
834	0.0194555500004918\\
835	0.0195223500004926\\
836	0.0195893500004934\\
837	0.0196565500004942\\
838	0.0197239500004949\\
839	0.0197915500004957\\
840	0.0198593500004965\\
841	0.0199273500004972\\
842	0.019995550000498\\
843	0.0200639500004987\\
844	0.0201325500004995\\
845	0.0202013500005002\\
846	0.0202703500005009\\
847	0.0203395500005016\\
848	0.0204089500005024\\
849	0.0204785500005031\\
850	0.0205483500005038\\
851	0.0206183500005045\\
852	0.0206885500005052\\
853	0.0207589500005059\\
854	0.0208295500005066\\
855	0.0209003500005073\\
856	0.020971350000508\\
857	0.0210425500005087\\
858	0.0211139500005093\\
859	0.02118555000051\\
860	0.0212573500005107\\
861	0.0213293500005113\\
862	0.021401550000512\\
863	0.0214739500005127\\
864	0.0215465500005133\\
865	0.0216193500005139\\
866	0.0216923500005146\\
867	0.0217655500005152\\
868	0.0218389500005158\\
869	0.0219125500005165\\
870	0.0219863500005171\\
871	0.0220603500005177\\
872	0.0221345500005183\\
873	0.0222089500005189\\
874	0.0222835500005195\\
875	0.0223583500005201\\
876	0.0224333500005207\\
877	0.0225085500005213\\
878	0.0225839500005218\\
879	0.0226595500005224\\
880	0.022735350000523\\
881	0.0228113500005236\\
882	0.0228875500005241\\
883	0.0229639500005247\\
884	0.0230405500005252\\
885	0.0231173500005258\\
886	0.0231943500005263\\
887	0.0232715500005268\\
888	0.0233489500005274\\
889	0.0234265500005279\\
890	0.0235043500005284\\
891	0.0235823500005289\\
892	0.0236605500005295\\
893	0.02373895000053\\
894	0.0238175500005305\\
895	0.023896350000531\\
896	0.0239753500005314\\
897	0.0240545500005319\\
898	0.0241339500005324\\
899	0.0242135500005329\\
900	0.0242933500005334\\
901	0.0243733500005338\\
902	0.0244535500005343\\
903	0.0245339500005348\\
904	0.0246145500005352\\
905	0.0246953500005357\\
906	0.0247763500005361\\
907	0.0248575500005366\\
908	0.024938950000537\\
909	0.0250205500005374\\
910	0.0251023500005379\\
911	0.0251843500005383\\
912	0.0252665500005387\\
913	0.0253489500005391\\
914	0.0254315500005395\\
915	0.0255143500005399\\
916	0.0255973500005403\\
917	0.0256805500005407\\
918	0.0257639500005411\\
919	0.0258475500005415\\
920	0.0259313500005419\\
921	0.0260153500005422\\
922	0.0260995500005426\\
923	0.026183950000543\\
924	0.0262685500005433\\
925	0.0263533500005437\\
926	0.0264383500005441\\
927	0.0265235500005444\\
928	0.0266089500005447\\
929	0.0266945500005451\\
930	0.0267803500005454\\
931	0.0268663500005458\\
932	0.0269525500005461\\
933	0.0270389500005464\\
934	0.0271255500005467\\
935	0.027212350000547\\
936	0.0272993500005473\\
937	0.0273865500005476\\
938	0.0274739500005479\\
939	0.0275615500005482\\
940	0.0276493500005485\\
941	0.0277373500005488\\
942	0.0278255500005491\\
943	0.0279139500005493\\
944	0.0280025500005496\\
945	0.0280913500005499\\
946	0.0281803500005501\\
947	0.0282695500005504\\
948	0.0283589500005506\\
949	0.0284485500005509\\
950	0.0285383500005511\\
951	0.0286283500005513\\
952	0.0287185500005516\\
953	0.0288089500005518\\
954	0.028899550000552\\
955	0.0289903500005522\\
956	0.0290813500005524\\
957	0.0291725500005527\\
958	0.0292639500005528\\
959	0.029355550000553\\
960	0.0294473500005532\\
961	0.0295393500005534\\
962	0.0296315500005536\\
963	0.0297239500005538\\
964	0.029816550000554\\
965	0.0299093500005541\\
966	0.0300023500005543\\
967	0.0300955500005545\\
968	0.0301889500005546\\
969	0.0302825500005548\\
970	0.0303763500005549\\
971	0.030470350000555\\
972	0.0305645500005552\\
973	0.0306589500005553\\
974	0.0307535500005554\\
975	0.0308483500005556\\
976	0.0309433500005557\\
977	0.0310385500005558\\
978	0.0311339500005559\\
979	0.031229550000556\\
980	0.0313253500005561\\
981	0.0314213500005562\\
982	0.0315175500005563\\
983	0.0316139500005563\\
984	0.0317105500005564\\
985	0.0318073500005565\\
986	0.0319043500005566\\
987	0.0320015500005566\\
988	0.0320989500005567\\
989	0.0321965500005568\\
990	0.0322943500005568\\
991	0.0323923500005569\\
992	0.0324905500005569\\
993	0.0325889500005569\\
994	0.032687550000557\\
995	0.032786350000557\\
996	0.032885350000557\\
997	0.032984550000557\\
998	0.0330839500005571\\
999	0.0331835500005571\\
1000	0.0332833500005571\\
};
\addlegendentry{$N=1000$}

\addplot [color=black, dashed]
  table[row sep=crcr]{%
1	0.00328349999999996\\
2	0.00318549999999996\\
3	0.00308949999999997\\
4	0.00299549999999997\\
5	0.00290349999999997\\
6	0.00281349999999997\\
7	0.00272549999999997\\
8	0.00263949999999997\\
9	0.00255549999999997\\
10	0.00247349999999997\\
11	0.00239349999999997\\
12	0.00231549999999997\\
13	0.00223949999999997\\
14	0.00216549999999997\\
15	0.00209349999999997\\
16	0.00202349999999997\\
17	0.00195549999999998\\
18	0.00188949999999998\\
19	0.00182549999999998\\
20	0.00176349999999998\\
21	0.00170349999999998\\
22	0.00164549999999998\\
23	0.00158949999999998\\
24	0.00153549999999998\\
25	0.00148349999999998\\
26	0.00143349999999998\\
27	0.00138549999999998\\
28	0.00133949999999998\\
29	0.00129549999999998\\
30	0.00125349999999998\\
31	0.00121349999999998\\
32	0.00117549999999998\\
33	0.00113949999999998\\
34	0.00110549999999998\\
35	0.00107349999999998\\
36	0.00104349999999998\\
37	0.00101549999999998\\
38	0.000989499999999981\\
39	0.000965499999999981\\
40	0.000943499999999982\\
41	0.000923499999999983\\
42	0.000905499999999983\\
43	0.000889499999999984\\
44	0.000875499999999985\\
45	0.000863499999999986\\
46	0.000853499999999987\\
47	0.000845499999999988\\
48	0.000839499999999989\\
49	0.00083549999999999\\
50	0.000833499999999991\\
51	0.000833499999999992\\
52	0.000835499999999993\\
53	0.000839499999999994\\
54	0.000845499999999996\\
55	0.000853499999999997\\
56	0.000863499999999998\\
57	0.0008755\\
58	0.000889500000000001\\
59	0.000905500000000003\\
60	0.000923500000000005\\
61	0.000943500000000007\\
62	0.000965500000000009\\
63	0.000989500000000011\\
64	0.00101550000000001\\
65	0.00104350000000001\\
66	0.00107350000000002\\
67	0.00110550000000002\\
68	0.00113950000000002\\
69	0.00117550000000002\\
70	0.00121350000000003\\
71	0.00125350000000003\\
72	0.00129550000000003\\
73	0.00133950000000003\\
74	0.00138550000000004\\
75	0.00143350000000004\\
76	0.00148350000000004\\
77	0.00153550000000004\\
78	0.00158950000000004\\
79	0.00164550000000005\\
80	0.00170350000000005\\
81	0.00176350000000005\\
82	0.00182550000000005\\
83	0.00188950000000005\\
84	0.00195550000000006\\
85	0.00202350000000006\\
86	0.00209350000000006\\
87	0.00216550000000006\\
88	0.00223950000000006\\
89	0.00231550000000006\\
90	0.00239350000000006\\
91	0.00247350000000006\\
92	0.00255550000000006\\
93	0.00263950000000006\\
94	0.00272550000000007\\
95	0.00281350000000007\\
96	0.00290350000000007\\
97	0.00299550000000007\\
98	0.00308950000000007\\
99	0.00318550000000007\\
100	0.00328350000000007\\
};
\addlegendentry{$N=100$}

\addplot [color=red, dashdotted]
  table[row sep=crcr]{%
1	0.000285\\
2	0.000205\\
3	0.000145\\
4	0.000105\\
5	8.49999999999999e-05\\
6	8.50000000000001e-05\\
7	0.000105\\
8	0.000145\\
9	0.000205\\
10	0.000285\\
};
\addlegendentry{$N=10$}

\end{axis}
\end{tikzpicture}%
  }}
  \caption{$\var{\hat\phi_n}$ as function of antenna index $n$, for
    different total numbers of antennas, $N$, for the line (radio
    stripe) topology in Figure~\ref{fig:lrs}, for ${\bQ
      =10^{-4}\cdot\bI}$.  Note the logarithmic scale; the minimum
    occurs when $n=N/2$.\label{fig:linevar}}
\end{figure}

Figure~\ref{fig:linevar} shows $\var{\hat\phi_n}$ as function of $n$,
for some different values of $N$.  The largest variances occur at the
ends of the stripe.  When $N\to\infty$, {the smallest} among
$\{\var{\hat\phi_n}\}$ (which, as seen in the figure, is in the middle
of the stripe) -- and therefore all of them -- grow without bound.

We remind the reader   about the point made earlier regarding arithmetic
$\mtp$. Also, we stress that the inequality (\ref{eq:varbound})
is not tight at all; its only purpose is to show the unboundedness.

\subsection{Ring  Topology}

\begin{figure}[t!]
  \begin{center}
    \begin{tikzpicture}[xscale=1.3,yscale=.9]
      \node (a1) at (0,0) [draw,rectangle,thick,minimum width=.7cm,minimum height=.5cm] {$1$};
      \node (a2) at (1,0) [draw,rectangle,thick,minimum width=.7cm,minimum height=.5cm] {$2$};
      \node (a3) at (2,0) [draw,rectangle,thick,minimum width=.7cm,minimum height=.5cm] {$3$};
      \node (a4) at (3,0) [draw,rectangle,thick,minimum width=.7cm,minimum height=.5cm] {$4$};
      \node (a5) at (3,1) [draw,rectangle,thick,minimum width=.7cm,minimum height=.5cm] {$5$};
      \node (a7) at (3,2) [draw,rectangle,thick,minimum width=.7cm,minimum height=.5cm] {$6$};
      \node (a8) at (2,2) [draw,rectangle,thick,minimum width=.7cm,minimum height=.5cm] {$7$};
      \node (a9) at (1,2) [draw,rectangle,thick,minimum width=.7cm,minimum height=.5cm] {$8$};
      \node (a10) at (0,2) [draw,rectangle,thick,minimum width=.7cm,minimum height=.5cm] {$\cdots$};
      \node (a12) at (0,1) [draw,rectangle,thick,minimum width=.7cm,minimum height=.5cm] {$N$};
      
      \draw [thick] (a1)   -- (a2);
      \draw [thick] (a2)   -- (a3);
      \draw [thick] (a3)   -- (a4);
      \draw [thick] (a4)   -- (a5);
      \draw [thick] (a5)   -- (a7);
      \draw [thick] (a7)   -- (a8);
      \draw [thick] (a8)   -- (a9);
      \draw [thick] (a9)   -- (a10);
      \draw [thick] (a10)   -- (a12);
      \draw [thick] (a12)   -- (a1);
      
    \end{tikzpicture}
  \end{center}
  \caption{Ring topology, where every antenna performs measurements on
    its two neighbors.\label{fig:2}}
\end{figure}
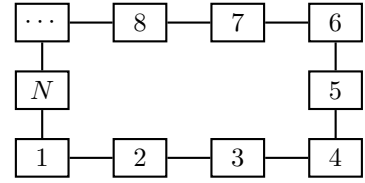

A variation on the previous example is a ring topology, obtained from
the line topology by connecting nodes $1$ and $N$
(Figure~\ref{fig:2}).  The Laplacian becomes a circulant matrix, which
can be diagonalized using the discrete Fourier transform
\cite{gray2006toeplitz}. Because of the symmetry, the eigenvalues of
$\bL$ appear in pairs and the eigenvectors appear in
complex-conjugated pairs, such that appropriate linear combinations
thereof yield real-valued eigenvectors. Among these eigenvectors, one
is
\begin{align}
\bz = \frac{1}{\sqrt{N}} \begin{bmatrix} 1 \\ \cos(2\pi/N) \\ \cos(2\cdot 2\pi/N) \\ \vdots \\ \cos((N-1)2\pi/N) \end{bmatrix},
\end{align} 
which has unit norm, $\Vert\bz\Vert=1$, and corresponding eigenvalue
$4\sin^2(\pi/N)$.  (One can alternatively find this eigenvector from
results in \cite{cvetkovic1980spectra}.)

In this example all nodes are statistically identical and a
calculation somewhat similar to, but simpler than, the one in
Section~\ref{sec:stripe} (we leave the details to the reader) shows
that for any $n$,
\begin{align}
\var{\hat\phi_n} \ge  { \frac{1/N}{4\sin^2(\pi/N)}} \ge  { \frac{1/N}{4 \pi^2/N^2}} =  \frac{N}{4\pi^2}.
\end{align}
Thus as $N\to\infty$ the ring exhibits the same behavior as the line
topology: the variances grow without bound.

\subsection{Large-Intelligent-Surface Topology}\label{sec:lis2d}

In the next example we consider a large intelligent
surface,\footnote{A large intelligent surface is not to be confused
with a \emph{reflecting} intelligent surface (RIS), also known as an
intelligent reflecting surface (IRS).  The use of a RIS (IRS) entails
its own calibration challenges \cite{zhang2023over}, but they are of a
totally different kind and unrelated to the material presented
herein.} envisioned by many as a main physical-layer solution for 6G
\cite{hu2018beyond,van2019radioweaves}. The idea is to cover a large
surface (for example, a wall or ceiling of a building) by antennas
that may be, for aesthetic reasons, physically integrated into the
surface itself.

\begin{figure}[t!]
  \begin{center}
    \begin{tikzpicture}[xscale=1.3,yscale=0.9]
      \draw[decorate,dashed, thick,decoration={random steps,segment length=1.5pt,amplitude=.8pt}] (2.5,2.5) rectangle (-0.5,-0.5)
            node at (-0.5,1) [anchor=east] {$\Omega$};
            
            \node (a1) at (0,0) [draw,rectangle,thick,minimum width=.7cm,minimum height=.5cm] {$1$};
            \node (a2) at (1,0) [draw,rectangle,thick,minimum width=.7cm,minimum height=.5cm] {$2$};
            \node (a3) at (2,0) [draw,rectangle,thick,minimum width=.7cm,minimum height=.5cm] {$3$};
            \node (a4) at (3,0) [draw,rectangle,thick,minimum width=.7cm,minimum height=.5cm] {$\cdots$};
            \node (a5) at (0,1) [draw,rectangle,thick,minimum width=.7cm,minimum height=.5cm] {$4$};
            \node (a6) at (1,1) [draw,rectangle,thick,minimum width=.7cm,minimum height=.5cm] {$5$};
            \node (a7) at (2,1) [draw,rectangle,thick,minimum width=.7cm,minimum height=.5cm] {$6$};
            \node (a8) at (3,1) [draw,rectangle,thick,minimum width=.7cm,minimum height=.5cm] {$\cdots$};
            \node (a9) at (0,2) [draw,rectangle,thick,minimum width=.7cm,minimum height=.5cm] {$7$};
            \node (a10) at (1,2) [draw,rectangle,thick,minimum width=.7cm,minimum height=.5cm] {$8$};
            \node (a11) at (2,2) [draw,rectangle,thick,minimum width=.7cm,minimum height=.5cm] {$\cdots$};
            \node (a12) at (3,2) [draw,rectangle,thick,minimum width=.7cm,minimum height=.5cm] {$\cdots$};
            \node (a13) at (0,3) [draw,rectangle,thick,minimum width=.7cm,minimum height=.5cm] {$\cdots$};
            \node (a14) at (1,3) [draw,rectangle,thick,minimum width=.7cm,minimum height=.5cm] {$\cdots$};
            \node (a15) at (2,3) [draw,rectangle,thick,minimum width=.7cm,minimum height=.5cm] {$\cdots$};
            \node (a16) at (3,3) [draw,rectangle,thick,minimum width=.7cm,minimum height=.5cm] {$N$};
            
            \draw [thick] (a1)   -- (a2);
            \draw [thick] (a2)   -- (a3);
            \draw [thick] (a3)   -- (a4);
            \draw [thick] (a5)   -- (a6);
            \draw [thick] (a6)   -- (a7);
            \draw [thick] (a7)   -- (a8);
            \draw [thick] (a9)   -- (a10);
            \draw [thick] (a10)   -- (a11);
            \draw [thick] (a11)   -- (a12);
            \draw [thick] (a13)   -- (a14);
            \draw [thick] (a14)   -- (a15);
            \draw [thick] (a15)   -- (a16);
            \draw [thick] (a1)   -- (a5);
            \draw [thick] (a2)   -- (a6);
            \draw [thick] (a3)   -- (a7);
            \draw [thick] (a4)   -- (a8);
            \draw [thick] (a5)   -- (a9);
            \draw [thick] (a6)   -- (a10);
            \draw [thick] (a7)   -- (a11);
            \draw [thick] (a8)   -- (a12);
            \draw [thick] (a9)   -- (a13);
            \draw [thick] (a10)   -- (a14);
            \draw [thick] (a11)   -- (a15);
            \draw [thick] (a12)   -- (a16);
            \draw [thick] (a1)   -- (a6);
            \draw [thick] (a5)   -- (a10);
            \draw [thick] (a9)   -- (a14);
            \draw [thick] (a2)   -- (a7);
            \draw [thick] (a3)   -- (a8);
            \draw [thick] (a11)   -- (a16);
            \draw [thick] (a10)   -- (a15);
            \draw [thick] (a7)   -- (a12);
            \draw [thick] (a6)   -- (a11);
            \draw [thick] (a5)   -- (a2);
            \draw [thick] (a6)   -- (a3);
            \draw [thick] (a7)   -- (a4);
            \draw [thick] (a9)   -- (a6);
            \draw [thick] (a13)   -- (a10);
            \draw [thick] (a14)   -- (a11);
            \draw [thick] (a15)   -- (a12);
            \draw [thick] (a11)   -- (a8);
            \draw [thick] (a10)   -- (a7);
    \end{tikzpicture}
  \end{center}
\caption{Large-intelligent-surface topology, where each antenna
  performs measurements on its north-south, east-west,
  southeast-northwest and southwest-northeast closest
  neighbors.\label{fig:3}}
\end{figure}
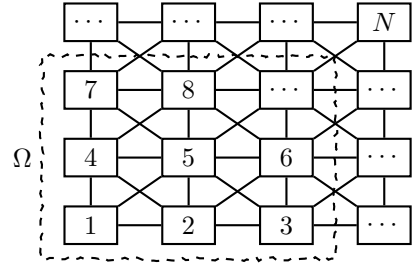

We assume, for the example, that each antenna can perform calibration
measurements on its eight closest neighbors (except for the antennas
on the border), as shown in Figure~\ref{fig:3}. The Laplacian is
easily written down but analysis in closed form, beyond the general
formulas already obtained, is cumbersome.

\begin{figure}[t!]
  \centerline{\scalebox{1}{
\begin{tikzpicture}

\begin{axis}[%
thick,
width=7.607cm,
height=4cm,
at={(0cm,0cm)},
scale only axis,
xmode=log,
xmin=5,
xmax=1000,
xminorticks=true,
xlabel style={font=\color{white!15!black}},
xlabel={number of antennas ($N$)},
ymin=2e-05,
ymax=8e-05,
ylabel style={font=\color{white!15!black}},
ylabel={Var$(\hat\phi_1)$},
axis background/.style={fill=white},
legend style={at={(0.03,0.97)}, anchor=north west, legend cell align=left, align=left, draw=white!15!black},
ylabel near ticks,
xlabel near ticks,
ylabel style={font=\small},
xlabel style={font=\small},
legend style={font=\small},,
ticklabel style={font=\small}
]
\addplot [color=blue]
  table[row sep=crcr]{%
9	3.02184235517569e-05\\
16	3.795549311095e-05\\
25	4.37197872132012e-05\\
36	4.82929649107675e-05\\
49	5.20748060548749e-05\\
100	6.05758864228085e-05\\
144	6.48107944804062e-05\\
225	6.99131602122236e-05\\
289	7.27422063044089e-05\\
400	7.63856542735077e-05\\
625	8.13414836810649e-05\\
900	8.53573750605105e-05\\
};
\addlegendentry{(i) using all $N$ antennas}

\addplot [color=black, dashed]
  table[row sep=crcr]{%
9	3.02184235517569e-05\\
16	3.02184235517569e-05\\
25	3.02184235517569e-05\\
36	3.02184235517569e-05\\
49	3.02184235517569e-05\\
100	3.02184235517569e-05\\
144	3.02184235517569e-05\\
225	3.02184235517569e-05\\
289	3.02184235517569e-05\\
400	3.02184235517569e-05\\
625	3.02184235517569e-05\\
900	3.02184235517569e-05\\
};
\addlegendentry{(ii) using antennas in $\Omega$}

\end{axis}
\end{tikzpicture}%

  }}
  \caption{For the topology in Figure \ref{fig:3}, and  ${\bQ =10^{-4}\cdot\bI}$: (i)
    $\var{\hat\phi_1}$ when using all available calibration
    measurements; (ii) $\var{\hat\phi_1}$ when using only the
    calibration measurements among antennas in $\Omega$, consisting of
    the nine antennas in the $3\times 3$ lower left corner of the
    surface.\label{fig:LISvar1}  }
\end{figure}
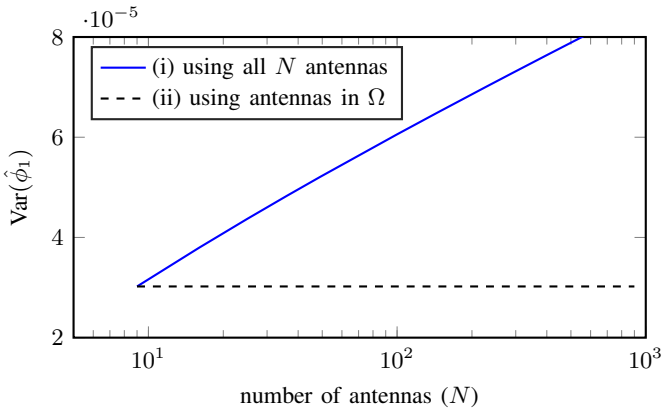

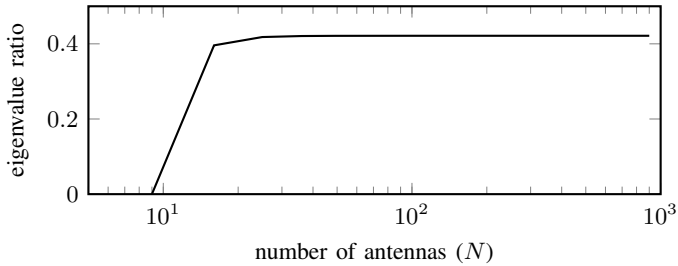
\begin{figure}[t!]
  \centerline{\scalebox{1}{
\begin{tikzpicture}

\begin{axis}[%
thick,
width=7.607cm,
height=2.5cm,
at={(0cm,0cm)},
scale only axis,
xmode=log,
xmin=5,
xmax=1000,
xminorticks=true,
xlabel style={font=\color{white!15!black}},
xlabel={number of antennas ($N$)},
ymin=0,
ymax=0.5,
ylabel style={font=\color{white!15!black}},
ylabel={eigenvalue ratio},
axis background/.style={fill=white},
ylabel near ticks,
xlabel near ticks,
ylabel style={font=\small},
xlabel style={font=\small},
legend style={font=\small},,
ticklabel style={font=\small}
]
\addplot [color=black, forget plot]
  table[row sep=crcr]{%
9	-9.99086708522359e-16\\
16	0.395976187995846\\
25	0.418051435705787\\
36	0.420761461922057\\
49	0.4212665059486\\
100	0.421438716431401\\
144	0.421445521447841\\
225	0.421447104173535\\
289	0.42144729228965\\
400	0.42144736848048\\
625	0.421447391118016\\
900	0.421447394454647\\
};
\end{axis}
\end{tikzpicture}%

  }}
  \caption{For the topology in Figure \ref{fig:3}:  the largest eigenvalue of $\bK_b-\bK_a$ relative to
    the largest eigenvalue of $\bK_b$.\label{fig:LISvar2}}
\end{figure}

Figure~\ref{fig:LISvar1} shows the following:
\begin{enumerate}[(i)]
\item the estimation error variance for the antenna in the lower-left
  corner (number 1 in Figure~\ref{fig:3}), $\var{\hat\phi_1}$, for
  case (a), when using all $M$ calibration measurements, and

\item $\var{\hat\phi_1}$ for case (b) when using only the $M_\Omega$
  calibration measurements among antennas in a subset $\Omega$,
  consisting of the nine antennas in the $3\times 3$ lower left corner
  of the surface.
\end{enumerate}
We make the following  observations:
\begin{itemize}
\item $\var{\hat\phi_1}$ in case (a) increases when using more
  antennas, though not as fast as for the line topology
  (cf.~(\ref{eq:varbound})).  We conjecture that the variance is
  unbounded when $N\to \infty$.

\item $\var{\hat\phi_1}$ in case (b) is, of course, independent of
  $N$.
\end{itemize}

Next, Figure~\ref{fig:LISvar2} shows the largest possible improvement
in beamforming accuracy when using all antennas for calibration,
compared to when using only the antennas in $\Omega$, quantified via
the largest eigenvalue of $\bK_b-\bK_a$ relative to the largest
eigenvalue of $\bK_b$.  This is the largest value that $V_b-V_a$ can
attain relative to the largest possible value that $V_b$ can attain,
for any unit-norm $\bv$. (Note that ${\Vert\bv\Vert^2 =
  \bv^H\bv=\bp^H\bZom^T\bZom\bp=\bp^H\bp=\Vert\bp\Vert^2}$.)  A
significant improvement in performance when beamforming from antennas
in $\Omega$ can be achieved by using calibration measurements from
antennas outside of $\Omega$. But the gain levels off quickly. Beyond
$N=16$ the curve is almost flat so there is no point in practice in
going beyond the $4\times 4$ lower-left corner.

\subsection{Complete Graph Topology}\label{sec:complete}

We end with an exposition of the situation when all $N$ antennas
measure on all others.  In this case, $\mG$ is a complete graph with
$M=N(N-1)/2$ edges. The Laplacian is
\begin{align}
  \bL = N \bI - \bu\bu^T.
\end{align}
Recalling that $\bu^T\bZ=\bzero$, we find that 
\begin{align}\label{eq:covcomplete}
  \cov{\hat\bphi}=\bZ(\bZ^T\bL\bZ)^{-1}\bZ^T = \frac{1}{N} \bZ \bZ^T.
\end{align}
The variance of $\hat\phi_n$, for an arbitrary $n$, follows as,
\begin{align}
  \var{\hat\phi_n} = &
  \be_n^T \cov{\hat\bphi}\be_n= \frac{1}{N} \be_n^T\bZ \bZ^T\be_n \nonumber \\
  = & \frac{1}{N} \be_n^T\pp{\bI-\frac{1}{N}\bu\bu^T}\be_n \nonumber \\
  = & \frac{1}{N} \pp{1 - \frac{1}{N}} = \frac{1}{N} - \frac{1}{N^2}.
\end{align}
When adding more and more antennas, this variance goes to zero:
$\var{\hat\phi_n} \to 0$ as $N\to\infty$, uniformly over $n$.  This is
a consequence of the dense topology, and the conclusion is the
opposite of that for the other topologies discussed above.  In the
present example, the number of unknowns is proportional to $N$, but
the number of measurements is proportional to $N^2$, making the
problem more and more well-conditioned as $N$ increases.

\section{Massive Synchrony}\label{sec:massivesync}

Returning to question Q3 asked in Section~\ref{sec:preview}, what
topologies enable {massive synchrony} in the sense that $\{ \var{\hat
  \phi_n} \}$ remain bounded, or even vanish, when $N\to\infty$?  For
the complete graph (Section~\ref{sec:complete}), all variances tend to
zero. But for the line topology, in contrast
(Section~\ref{sec:stripe}) all variances tend to infinity.  It appears
plausible that the complete graph could be ``thinned out'' quite
substantially before encountering this phenomenon, but it is unclear
exactly how much.

One observation is that the  {average} variance can be
written in  terms of the Laplacian eigenvalues:
\begin{align}
  \frac{1}{N} \sum_{n=1}^N \var{\hat\phi_n} & =
  \frac{1}{N} \tr{\cov{ \hat\bphi}} =  \frac{1}{N} \tr{\bZ \bLam^{-1} \bZ^T  } \nonumber \\
  & =  \frac{1}{N} \tr{ \bLam^{-1} \bZ^T  \bZ } =  \frac{1}{N} \tr{ \bLam^{-1}   }  \nonumber \\
  & =  \frac{1}{N} \sum_{n=1}^{N-1} \frac{1}{[\bLam]_{nn}} .  \label{eq:meanbound}
\end{align}
For a {tree} topology, and for uncorrelated, homoskedastic noise
($\bQ=\bI$) one can use \cite[Theorem 4.3]{mohar1991eigenvalues} (see
also \cite{de2007old}) to write the right hand side of
(\ref{eq:meanbound}) as a constant times the {average distance}
between pairs of nodes in $\mG$.  For an arbitrary topology,
monotonicity (Section~\ref{sec:mono}) then gives an upper bound on the
average variance in terms of the average distance of a spanning tree.

Whether stronger results can be established remains open.  These
questions might have limited (at best) impact on the operation of
distributed antenna systems in practice, but come across as
interesting basic research problems.

\section{Concluding Remarks}

For R-calibration, it is enough to compute {a single} set of
calibration coefficients for the entire network, based on appropriate
bidirectional measurements between service antennas, and then use
these coefficients for all beamforming activities in the network.
This is so {despite the fact} that the phase estimation errors grow,
in some cases without bound, the more antennas are involved in the
calibration process.  An important consequence is that there is no
need to compute calibration coefficients associated with specific
``local areas'' or specific users. In fact, irrespective of the
network size, solving a single global calibration problem, and using
the so-obtained phase corrections everywhere in the network, is
optimal.

Extensions of the analyses may be possible. For example, one could
potentially work directly on $\{T_n,R_n\}$ to circumvent the
assumption on ``small'' errors induced by the $\mtp$ arithmetic,
although this appears very difficult due to the nonlinear nature of
the ensuing estimation problem.  Also, a complete characterization of
topologies for which $\{\var{\hat\phi_n}\}$ are bounded, or vanish,
with increasing $N$, is an open problem.  

The beamforming gain
analysis herein applies to reciprocity-based beamforming.
Corresponding analyses for other applications, such as sensing and
positioning, and for F-calibration, could be of interest too.  
An additional possible topic for future
work is to consider the overhead incurred by calibration measurements,
and its effect on the resulting spectral efficiency when the
antennas are used for multiuser beamforming.
 
\section*{Acknowledgment}

The author thanks the colleagues in the REINDEER project for extensive
discussions on antenna calibration, and Prof. Liesbet Van der Perre
for her useful comments on an early draft of the manuscript.

\end{document}